\documentclass[letterpaper,journal]{IEEEtran}

\usepackage{amsmath,amsfonts}
\usepackage{algorithmic}
\usepackage{algorithm}
\usepackage{array}
\usepackage[caption=false,font=normalsize,labelfont=sf,textfont=sf]{subfig}
\usepackage{textcomp}
\usepackage{stfloats}
\usepackage{url}
\usepackage{verbatim}
\usepackage{graphicx}
\usepackage{cite}
\usepackage{booktabs}
\usepackage{multirow}
\usepackage{xcolor}

\makeatletter
\def\@IEEENORMtitlevspace{1\baselineskip}
\def\@IEEEMINtitlevspace{0.75\baselineskip}
\makeatother
\renewcommand{\footnoterule}{%
  \kern-3pt
  \hrule width 0.4\columnwidth height 0.4pt
  \kern2.6pt}

\newcommand{\methodname}{SonicRender}
\newcommand{\tasg}{TASG}

\begin{document}
\pagestyle{plain}
\title{SonicRender: Structured and Time-Grounded Rendering of Multi-Event Audio Scene Generation}
\author{Yunrui Cai\textsuperscript{1,2,*},
Xu Li\textsuperscript{1,\textdagger},
Yan Rong\textsuperscript{1},
Dingdong Wang\textsuperscript{2},
Yucheng Zhou\textsuperscript{1},\\
Xixin Wu\textsuperscript{2},
Chen Zhang\textsuperscript{1},
Zhiyong Wu\textsuperscript{3},
Helen Meng\textsuperscript{2,\textdagger}\\[0.2em]
\parbox[t]{\textwidth}{\centering\small
\textsuperscript{1}Kling AI\\[0.1em]
\textsuperscript{2}
The Chinese University of Hong Kong\\[0.1em]
\textsuperscript{3}Shenzhen International Graduate School, Tsinghua University\\[0.2em]
\texttt{\{yrcai, hmmeng\}@se.cuhk.edu.hk, lixu15@kuaishou.com}}%
\thanks{\textsuperscript{*}Work done during Yunrui Cai's internship at Kling AI.}
\thanks{\textsuperscript{\textdagger}Corresponding authors: Xu Li and Helen Meng.}}
\maketitle
\thispagestyle{plain}

\begin{abstract}
While unified audio generators can synthesize speech, music, and sound effects within a single model, producing a plausible mixture is only part of the challenge: a complex scene request also specifies \emph{which} events occur, \emph{who} produces them, and \emph{when} they are heard. To satisfy these compositional constraints in a single generation pass, we present \methodname{}, a unified multi-event audio renderer.
% Generating a convincing mixture is only part of the challenge in unified audio generation: a complex request also specifies \emph{which} events should occur, \emph{who} produces them, and \emph{when} they should be heard. In this paper we present \methodname{}, a unified multi-event audio renderer that brings these constraints into a single generation process. 
Its Temporal Audio Scene Graph (\tasg{}) binds persistent source identities to localized speech, music, ambience, and sound-effect events, together with their content, acoustic roles, relations, and timestamped windows. We further introduce AbsTime, which supplies the audio latent stream with absolute-frame coordinates and valid-canvas duration. Through joint text--audio attention and conditional flow matching, the renderer learns to realize structured, timed scenes as coherent mixtures. On the MMAG timestamp benchmark, \methodname{} achieves speech/sound F1 scores of 0.79/0.60, exceeding the strongest reported scores in our comparison. On our SceneMix evaluation, it attains 85.4\% all-event recall, 83.9\% window-hit rate, and 95.7\% speaker accuracy. Subjective study further demonstrates significant improvements over all five comparison systems in both perceived audio quality and scene realization. Results across speech, general audio, music, multi-event load, and temporal-grounding analyses establish precise, composition-aware audio rendering within a unified generator. Demo page: \url{https://caiyunrui.github.io/SonicRender}.
\end{abstract}


\section{Introduction}
\label{sec:intro}

% \IEEEPARstart{A}{n} audio scene is more than a collection of sounds. A single request can specify dialogue between recurring characters, music that continues beneath the conversation, a brief effect that interrupts one turn, and the precise moments at which these activities begin. Rendering it requires control over which events occur, who produces them, how they coexist, and when they are heard. Unified models synthesize speech, music, environmental sound, and mixtures with shared architectures~\cite{qiang2026unisonate,mei2026dasheng,sun2026midasheng,dai2026qwenaudio}. We target \emph{event-complete, source-consistent, and time-grounded rendering of a multi-event scene in one generated waveform}.
\IEEEPARstart{R}{ecent} unified audio generators synthesize speech, music, environmental sound, and their mixtures within shared architectures~\cite{qiang2026unisonate,mei2026dasheng,sun2026midasheng,dai2026qwenaudio}. Yet rendering a realistic audio scene requires more than blending acoustic textures: a single request can specify dialogue between recurring characters, music that continues beneath the conversation, a brief sound effect that interrupts one turn, and the exact moments at which each activity occurs. Realizing such scenes demands joint control over which events occur, who produces them, how they coexist, and when they are heard. We target \emph{event-complete}, \emph{source-consistent}, and \emph{time-grounded} rendering of multi-event scenes in one generated waveform.

Even when a generated mixture sounds globally plausible, it can omit a quiet sound effect, assign words to the wrong speaker, truncate a sustained background, or misplace an event in time. These failures compound rapidly as overlapping activities and recurring dialogue turns accumulate. Recent mixed-audio studies and benchmarks~\cite{mei2026dasheng,dai2026qwenaudio,zheng2026mmag} highlight event completeness, temporal ordering, and speaker consistency as critical bottlenecks beyond global audio quality. Achieving high compositional fidelity requires keeping individual source and event constraints identifiable throughout generation, while modeling their acoustic interactions within a single mixture rather than synthesizing isolated stems.

\begin{figure}[!t]
\centering
\includegraphics[width=\linewidth]{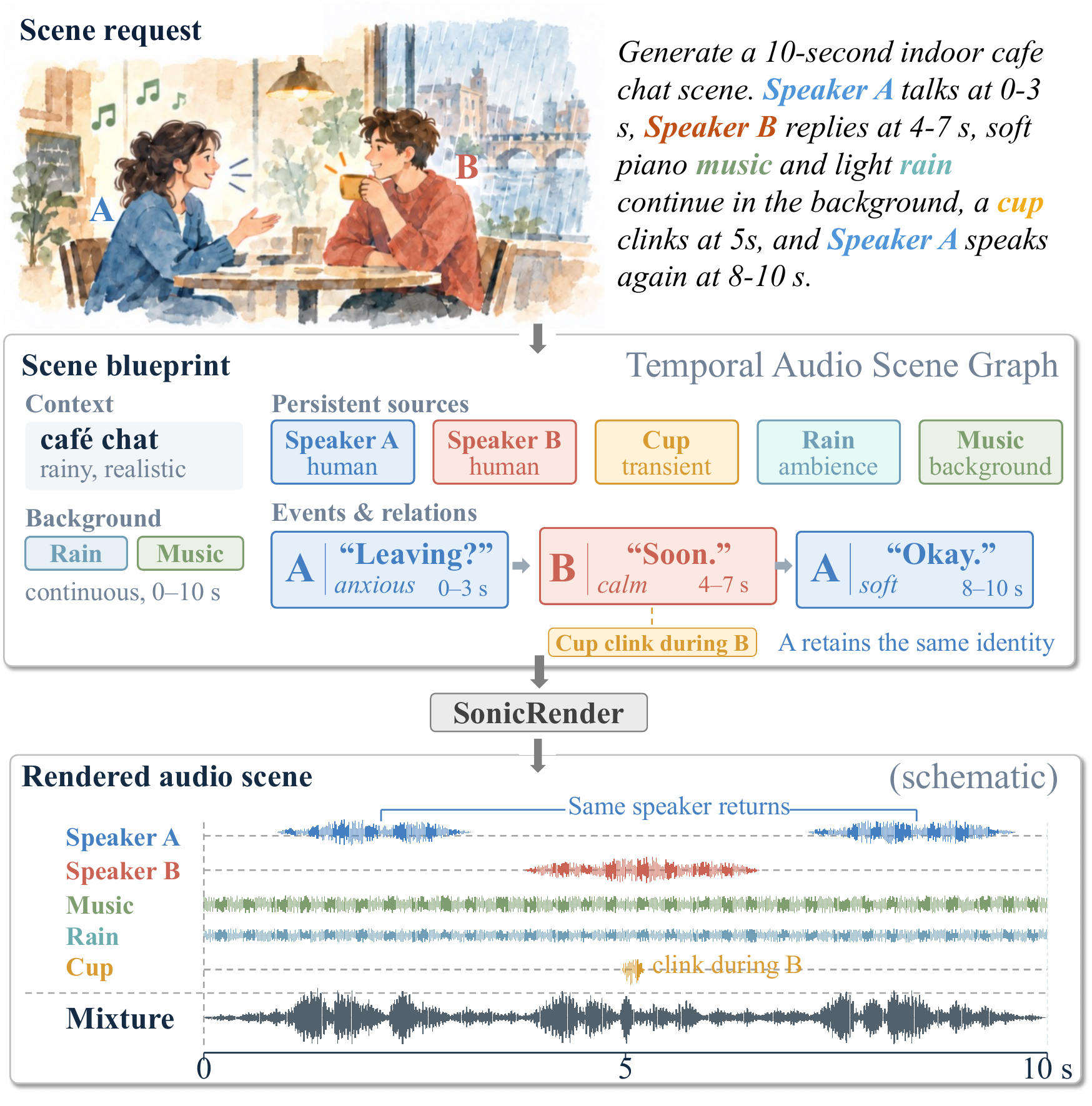}
\caption{Overview of SonicRender. A scene request is organized into a TASG blueprint that binds persistent sources to events, relations, and time windows. A unified renderer turns this blueprint into one coherent audio mixture with source continuity and event-level temporal control.}
\label{fig:teaser}
\end{figure}

% The first challenge is maintaining source--event bindings under composition. A source has a persistent acoustic identity; an event is a particular occurrence. A speaker may return with new words and delivery, while one music source continues across several turns. Structured conditions represent event order, temporal graphs, or role profiles~\cite{huang2023makeanaudio2,liu2026degdit,jiang2026controlaudio,dai2026qwenaudio}, but describing these elements is only part of the task: the renderer must jointly preserve identity, local content, acoustic role, and timing. We explicitly bind persistent sources to event instances and carry these references into a shared conditioning pathway.
The first challenge is maintaining source--event bindings under composition. A source carries a persistent acoustic identity, whereas an event is a localized occurrence: a speaker may return later with new words and delivery, while a music source persists across multiple dialogue turns. Existing structured conditions represent event order, temporal graphs, or role profiles~\cite{huang2023makeanaudio2,liu2026degdit,jiang2026controlaudio,dai2026qwenaudio}, yet flat or loosely linked descriptions still allow attributes to bleed across concurrent events. We bind persistent sources to localized events through explicit references: shared source records separate identity from turn-specific content, while event records and mixture relations keep brief effects and sustained backgrounds distinct under overlap. This organization directly targets the identity confusion, event omission, and layer truncation described above.

% The second challenge is grounding symbolic time in the generated audio. Existing approaches control timing through time matrices, structured text, and inference-time guidance~\cite{xie2024picoaudio,zheng2025picoaudio2,jiang2025freeaudio,zhang2026tag}. However, a timestamp such as ``6 seconds'' is a symbolic condition, whereas generation proceeds over latent frames. Sequence position alone does not explicitly communicate valid-canvas duration or establish a shared metric reference between these representations. A related script-to-output timing gap arises in audio--video generation~\cite{liu2026temporallink}. We seek a direct audio-side reference: a fixed physical second should correspond to the same latent coordinate as scene duration and event count vary.
The second challenge is grounding symbolic time in non-autoregressive latent generation. Existing approaches steer timing through time matrices, structured text, or inference-time guidance~\cite{xie2024picoaudio,zheng2025picoaudio2,jiang2025freeaudio,zhang2026tag}, and a similar script-to-output timing gap arises in audio--video generation~\cite{liu2026temporallink}. However, a textual timestamp such as ``6 seconds'' is a symbolic instruction, whereas a diffusion transformer (DiT) denoises a variable-length latent sequence in parallel. Relative sequence position alone (e.g., rotary embeddings) neither anchors a physical second to a fixed latent frame across varying audio durations nor communicates where the valid canvas ends. We introduce a direct audio-side metric reference so that any physical timestamp maps to a consistent latent coordinate regardless of scene duration or event count.

We introduce \methodname{}, a unified renderer combining source-aware specification with audio-side metric-time grounding. Its Temporal Audio Scene Graph (\tasg{}) binds persistent sources to local events, content, roles, relations, and metric windows in one serialized condition. AbsTime adds absolute-frame encodings to audio hidden states and valid-canvas duration to the global condition. A dual-stream multimodal diffusion transformer (MM-DiT)~\cite{esser2024mmdit} integrates event-level text through joint attention and a scene summary through global modulation. Trained with conditional flow matching~\cite{lipman2023flow}, shared-and-routed audio capacity renders speech, music, ambience, and effects as one mixture. Thus, \tasg{} specifies \emph{what} occurs and \emph{when}, while AbsTime grounds that time on the \emph{audio} timeline.

Public speech, general-audio, and music evaluations establish broad generation capability with a single checkpoint. Speech evaluations further cover Japanese, Korean, and Spanish, with additional languages, dialects, and accents showcased on our demo page. On MMAG~\cite{zheng2026mmag}, \methodname{} achieves speech/sound F1 scores of 0.79/0.60 against the original fractional-second ground-truth windows, leading the reported systems in our comparison. On 168 SceneMix requests, it reaches 85.4\% all-event recall, 83.9\% window-hit rate, and 95.7\% speaker accuracy, versus 74.6\%, 68.0\%, and 86.6\% for an information-matched natural-language variant. Event-load curves and matched ablations further establish the gains under denser composition and the role of audio-side coordinates in onset placement. A 15-listener study confirms that these objective gains are accompanied by significantly better audio quality and scene realization than all five comparison systems.

Our main contributions are as follows:
\begin{itemize}
\item We formulate \tasg{}, a source-aware and time-resolved conditioning representation that binds recurring acoustic identities to local events, content, roles, relations, and metric windows for joint multi-event rendering.
\item We introduce AbsTime, a lightweight module that grounds symbolic timestamps through absolute-frame coordinates and valid-canvas duration, offering a reusable design for temporal control in latent audio generators.
% \item We develop \methodname{}, a unified audio generation framework that combines joint text--audio attention, global scene modulation, and shared-and-routed acoustic capacity to generate speech, music, ambience, sound effects, and their complex mixtures with a single model.
\item We develop \methodname{}, a unified renderer that couples \tasg{} with joint text--audio attention for source--event interaction, global modulation for scene-wide context, and shared-and-routed acoustic capacity for heterogeneous overlapping events, generating speech, music, ambience, and sound effects within one mixture.
\item We introduce SceneMix, a compositional evaluation suite covering event completeness, recurring-speaker consistency, and temporal adherence in multi-event audio scenes. Evaluations on SceneMix and established benchmarks, together with controlled ablations and human listening tests, demonstrate the effectiveness of \methodname{}.
\end{itemize}

\section{Related Work}
\label{sec:related}

\subsection{Unified Audio and Multi-Source Scene Generation}
% AudioLDM~2~\cite{liu2024audioldm2}, UniSonate~\cite{qiang2026unisonate}, UniAudio~2.0~\cite{yang2026uniaudio2}, AudioCALM~\cite{liu2026audiocalm}, StepAudio~3~Gen~\cite{lin2026stepaudio3}, and AudioX~\cite{tian2026audiox} study shared representations and generation across acoustic domains or conditioning modalities. MiDashengLM-Gen~\cite{sun2026midasheng} combines autoregressive language modeling with flow-based generation. Broad acoustic coverage, however, does not by itself establish that every source and event in a mixed scene follows its individual instructions.
AudioLDM~2~\cite{liu2024audioldm2}, UniSonate~\cite{qiang2026unisonate}, UniAudio~2.0~\cite{yang2026uniaudio2}, AudioCALM~\cite{liu2026audiocalm}, StepAudio~3~Gen~\cite{lin2026stepaudio3}, and AudioX~\cite{tian2026audiox} study shared representations and generation across acoustic domains or conditioning modalities. MiDashengLM-Gen~\cite{sun2026midasheng} and Ming-Omni-TTS~\cite{ming2026omnittts} further unify speech, music, and sound generation within LLM-driven or omni-modal frameworks. Broad acoustic coverage, however, does not by itself establish that every source and event in a mixed scene follows its individual instructions.

Dasheng AudioGen~\cite{mei2026dasheng} uses structured multi-view captions and semantic--acoustic latents for native mixed-scene synthesis. SonicWeave~\cite{cai2026sonicweave} introduces chunk-routed MoE for heterogeneous, locally continuous acoustic content. We focus on finer control within such mixtures: realizing individually specified events and time windows while maintaining recurring identities and persistent backgrounds. Acoustic capacity and scene-level coherence address how a mixture sounds; source--event control additionally concerns whether each requested activity retains its content, ownership, and placement. This requires coupling source--event organization to a physical-time reference in the renderer.

VibeVoice~\cite{peng2026vibevoice} addresses extended conversational turn-taking, and ScenA~\cite{finkelson2026scena} conditions dialogue on multiple reference voices and acoustic context. Our text-specified scenes place speaker recurrence alongside music, ambience, and effects without mandatory voice references. WavJourney~\cite{liu2025wavjourney}, AuDirector~\cite{ren2026audirector}, and Audio-Oscar~\cite{duan2026audiooscar} instead coordinate specialist generation, planning, or post-production stages. We generate the complete mixture directly with one renderer.

While Qwen-Audio-3.0-Gen-Preview~\cite{dai2026qwenaudio} also incorporates role profiles and timeline formatting within a DiT backbone, it relies on text-side timing tokens without an explicit audio-side metric coordinate or valid-duration reference, and does not explicitly model cross-event acoustic roles (such as persistent background-to-foreground relations). \methodname{} couples a relational source--event scene graph with explicit audio-side frame and duration grounding, enabling strict completeness and temporal adherence in a single unified generation process.

% Qwen-Audio-3.0-Gen-Preview~\cite{dai2026qwenaudio} is closely related: it combines source/role profiles, linked dialogue turns, localized events, and rich-timeline training with a shared DiT and audio VAE. Our focus is the explicit coupling of source--event-addressable conditioning with audio-side metric-time grounding. \tasg{} organizes event requirements and relations; AbsTime supplies absolute frame coordinates and valid-canvas duration. Matched ablations isolate their effects on completeness and timestamp adherence.

\subsection{Structured Conditions and Source--Event Binding}
Structured conditions expose attributes that flat captions can entangle. Make-An-Audio~2~\cite{huang2023makeanaudio2} uses event-order planning; AudioComposer~\cite{wang2025audiocomposer} addresses content and style through fine-grained descriptions. DegDiT~\cite{liu2026degdit} embeds event semantics and temporal links with a graph transformer, while ControlAudio~\cite{jiang2026controlaudio} uses delimited event text and times with progressive conditioning. Qwen-Audio-3.0-Gen-Preview~\cite{dai2026qwenaudio} adopts a scene/role/event hierarchy. Beyond specifying these semantic units, direct mixture rendering requires their ownership and interactions to remain consistent across events.

% \tasg{} represents each persistent identity once and links it to separately timed events, preserving local content and acoustic roles within a common condition. A speaker can change delivery without changing identity, while a continuous background retains its role across foreground turns. Concurrent events remain separately addressable even when they overlap on the timeline. The text--audio transformer jointly processes these references and relations. Information-matched comparisons hold scene facts constant while changing their organization, isolating the effect of structured source--event organization. Structured music captions~\cite{wu2026scaps} also investigate representation format, but within music rather than heterogeneous source binding.
In contrast, \tasg{} decouples persistent source identities from localized event instances and links them via explicit reference IDs and mixture relations. This allows recurring speakers to vary their lexical content and delivery across turns while preserving vocal identity, and keeps overlapping foreground and background layers individually addressable by joint text--audio attention.

\subsection{Event-Level Temporal Control and Metric Grounding}
Stable Audio~\cite{evans2024timing} uses timing-conditioned latent diffusion; MC-Diffusion~\cite{guo2024mcdiffusion} adds timestamp and acoustic-style controls. PicoAudio~\cite{xie2024picoaudio} and PicoAudio2~\cite{zheng2025picoaudio2} steer events with temporally aligned data and time matrices. ControlAudio~\cite{jiang2026controlaudio} incorporates structured timing and phoneme conditions through progressive training and inference, whereas DegDiT~\cite{liu2026degdit} combines temporal attributes with event semantics and graph links. These approaches differ in how requested times enter the generation process.

Inference-time methods include Temporal Guidance Diffusion~\cite{du2024tgdiff}, which updates latents for desired event regions; FreeAudio~\cite{jiang2025freeaudio}, which combines timing plans, attention control, and latent composition; and TAG~\cite{zhang2026tag}, which guides attention with temporal descriptions. Audio ControlNet~\cite{zhu2026audiocontrolnet} supplies auxiliary local control features. In audiovisual generation, Temporal Context Routing~\cite{liu2026temporallink} addresses script-to-output timing through script-aware routing.
% while Soundwich~\cite{ning2026soundwich} produces independently controllable timed stems rather than a single mixture.

% \methodname{} grounds explicit \tasg{} windows through audio-side coordinates: AbsTime supplies absolute frame indices at a fixed latent rate and the valid-canvas duration. Specifying an event's requested interval and supplying the renderer's temporal reference serve different functions: the former describes the target, while the latter locates and delimits the generated sequence. This temporal reference is learned during end-to-end mixture training. Comparisons using identical structured inputs with and without AbsTime isolate its contribution through onset-position and tolerance analyses.
\methodname{} bridges symbolic text timestamps and parallel latent denoising through a lightweight AbsTime, without inference-time attention surgery or external time matrices. By injecting unnormalized frame indices at a fixed latent rate alongside valid-canvas duration, the renderer learns a direct correspondence between textual clock values and physical audio positions during end-to-end training.

\subsection{Evaluating Compositional and Temporal Fidelity}
Global audio--text similarity (e.g., CLAP) and distributional distances (e.g., FAD, KL) do not resolve whether individual source, event, and timing instructions are fulfilled. Recent benchmarks move toward finer-grained evaluation: AudioTime~\cite{xie2024audiotime} tests temporally aligned sound descriptions, while T2A-Feedback~\cite{wang2025t2afeedback} and instruction-following protocols~\cite{kuan2026instruction} examine event presence and ordering. MMAG~\cite{zheng2026mmag} extends evaluation to mixed audio across speech, music, sound effects, and timestamped segments.

However, existing benchmarks rarely stress-test dense multi-event compositions, nor do they evaluate text-prompted speaker identity consistency across recurring dialogue turns without voice references. To complement MMAG, we introduce SceneMix to jointly evaluate event recall, non-speech retention, recurring-speaker accuracy, and window adherence under increasing compositional load.

\section{Methodology}
\label{sec:method}

\subsection{Temporal Audio Scene Graph}
\label{sec:scene_spec}

We represent a requested acoustic scene by a structured blueprint
\begin{equation}
\mathcal{B}=(\mathcal{G},\mathcal{S},\mathcal{E},\mathcal{R},\mathcal{T}),
\label{eq:blueprint}
\end{equation}
where $\mathcal{G}$ describes global context, $\mathcal{S}$ contains persistent acoustic sources, $\mathcal{E}=\{E_k\}_{k=1}^{K}$ contains event instances, $\mathcal{R}$ specifies cross-event relations, and $\mathcal{T}=\{\tau_k\}_{k=1}^{K}$ collects metric time constraints. An event is defined as
\begin{equation}
E_k=(s_k,c_k,a_k,r_k,\tau_k),
\label{eq:event}
\end{equation}
where $s_k\in\mathcal{S}$ is its source identifier, $c_k$ its linguistic or semantic content, $a_k$ its local attributes, $r_k$ its acoustic role, and $\tau_k=[b_k,e_k]$ its requested temporal window when specified. Stable voice/timbre traits belong to the source record; language, emotion, delivery, temporal extent, and repetition belong to individual events. Thus, one speaker can change delivery across turns without becoming a new source, and a brief effect remains distinct from a persistent background. Relations specify acoustic interactions such as \emph{background to}, while event intervals also express ordering and overlap. $\mathcal{T}$ collects these intervals rather than duplicating them as a second schedule.

We call the serialized form Temporal Audio Scene Graph (\tasg{}). The canonical text orders scene context, source records, event records, and mixture relations. Field delimiters and identifiers remain in the text supplied to the encoder; the graph's connectivity is expressed through reusable references. Canonical formatting reduces variation in how equivalent scene requirements are expressed, while source--event factorization separates reusable identity information from event-specific content, delivery, and timing. Together, they simplify the associations the renderer must learn between scene instructions and acoustic realizations. A shortened example illustrates recurring-source binding:
\begin{quote}
\footnotesize
\begin{verbatim}
<types="speech,music", 2-speaker cafe chat/>
<source id="A" kind="speaker"/>
<source id="B" kind="speaker"/>
<source id="M" kind="music"/>
<event id="E1" source="A"
  time="[00:00 - 00:03]"
  type="speech">"Leaving?"</event>
<event id="E2" source="B"
  time="[00:04 - 00:07]"
  type="speech">"Soon."</event>
<event id="E3" source="A"
  time="[00:08 - 00:10]"
  type="speech">"Okay."</event>
<event id="E4" source="M"
  time="[00:00 - 00:10]"
  type="music" role="background"
  extent="persistent">soft piano</event>
<mix>E4 background_to E1 |
     E4 background_to E2 |
     E4 background_to E3</mix>
\end{verbatim}
\end{quote}
The repeated identifier \texttt{A} binds two speech turns to the same speaker, while \texttt{E1} and \texttt{E3} retain distinct content and timing. Speech and lyrics are stored verbatim; non-vocal events carry acoustic descriptions. This factorization makes both source continuity and event-local changes explicit within one condition. Timestamps use ordinary textual clock values, processed by the same tokenizer as the remaining fields. The renderer therefore consumes the structured scene directly through its text stream.

\begin{figure*}[t]
\centering
\includegraphics[width=0.92\textwidth]{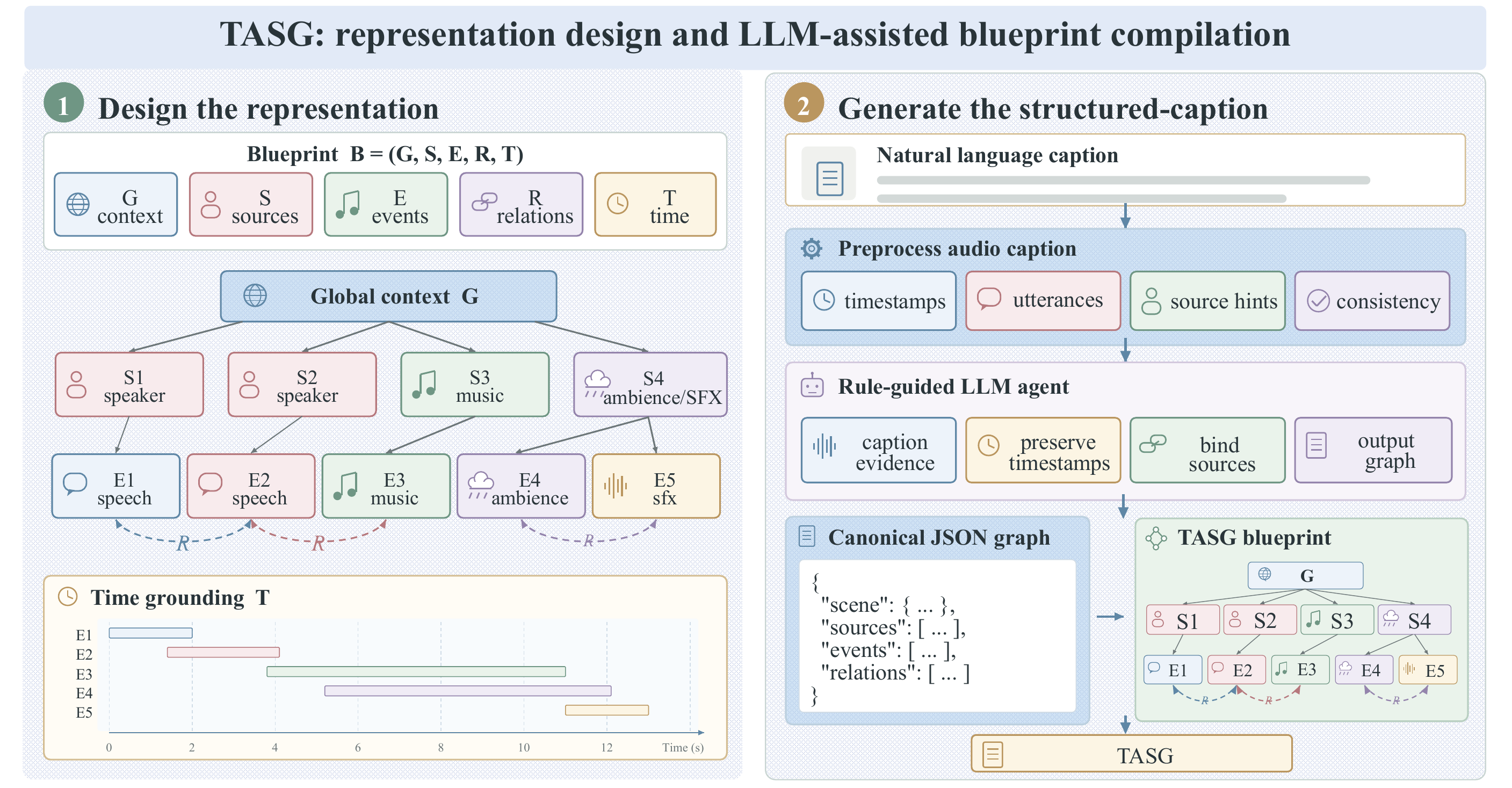}
\caption{TASG representation and LLM-assisted compilation. Left: global context, persistent sources, event instances, relations, and temporal windows define the scene blueprint. Right: a rule-guided LLM organizes a natural-language caption into a canonical scene graph for serialization as the timed generation condition.}
\label{fig:asg}
\end{figure*}

% For training-data annotation, we obtain the original natural-language audio descriptions using AudioMap~\cite{rong2026audiomap} and convert them into canonical \tasg{} records, preserving the described event content, source attributes, and timestamps.
For training-data annotation, we obtain time-aligned dense audio descriptions using AudioMap~\cite{rong2026audiomap}, which jointly captures timestamped speech utterances, speaker traits, and non-speech acoustic events, and then convert them into canonical \tasg{} records, preserving the described event content, source attributes, and timestamps.
For free-form input, a rule-guided language-model compiler can convert a caption into a canonical JSON scene graph, which is deterministically formatted into the compact \tasg{} tag sequence.
% : it identifies recurring sources, extracts event instances and their acoustic relationships, preserves quoted utterances and user-supplied timestamps, and constructs a consistent schedule for relative-only instructions. 
This provides one interface for explicit timed commands and ordinary scene requests. Renderer-level experiments directly use reviewed canonical specifications, so the experiment isolates the model's ability to realize the scene once its constraints are specified. Figure~\ref{fig:asg} illustrates the representation and compilation pathway.

During conditional training, the structured textual input and the audio mixture form a supervised pair. Source links and event intervals occur in the condition alongside the acoustic realization in the target mixture. This organization exposes recurring identity and local temporal attributes to the learning objective without demanding a source-separated waveform or an additional alignment network. 

\subsection{Audio-Side Metric-Time Grounding}
\label{sec:abstime}

In \tasg{}, the event window $\tau_k$ is expressed in physical seconds, while the audio generator processes a latent sequence. Let $f_s$ be the waveform sampling rate and $d$ the audio autoencoder downsampling factor. The latent rate is $r=f_s/d$, and a physical time $\tau$ corresponds approximately to audio frame $f(\tau)=r\tau$. AbsTime provides two complementary signals that make this metric relationship available to the audio computation.

Each valid latent frame $f\in\{0,\ldots,F_i-1\}$ is encoded by a sinusoidal positional map followed by a two-layer SiLU MLP, $a_f=\phi(f)$. The audio hidden state becomes
\begin{equation}
\tilde h_f=h_f+g_{\mathrm{abs}}\phi(f).
\label{eq:abstime}
\end{equation}
The sinusoidal map uses unit-scale frame indices, with frequencies 
\begin{equation}
\omega_j=10000^{-j/(H-1)} 
\end{equation}
and components $[\sin(f\omega_j),\cos(f\omega_j)]_{j=0}^{H-1}$, where $2H$ is the sinusoidal embedding dimension. The coordinate is absolute, rather than normalized by the canvas length $F_i$: at a fixed latent rate, the same physical time receives the same embedding across different durations. AbsTime supplements the backbone's rotary position encoding by placing an explicit frame reference in the audio hidden states, where it participates in attention values and subsequent feed-forward computation as well as queries and keys. It thus provides a direct metric-time signal throughout the audio pathway.

The number of valid audio latent frames $F_i$, obtained from the sample's valid mask, is encoded through a projection $\psi$ and added to the global diffusion-conditioning vector $q_i$:
\begin{equation}
\tilde q_i=q_i+g_{\mathrm{dur}}\psi(F_i).
\label{eq:duration}
\end{equation}
Here $q_i$ is the base flow-conditioning vector, and $\psi$ uses the same sinusoidal--MLP construction with separate parameters. The frame-wise pathway locates audio tokens; the global pathway communicates how far the valid timeline extends. The duration-enriched $\tilde q_i$ drives both text pooling and adaptive layer normalization in the renderer below. Both temporal signals remain active when semantic conditions are dropped for classifier-free guidance, so conditional and unconditional predictions refer to the same audio canvas. The scalar gates $g_{\mathrm{abs}}$ and $g_{\mathrm{dur}}$ start at zero, preserving the underlying renderer at initialization and enabling a gradual incorporation of metric-time information. Padding is excluded from attention and the training loss, and does not contribute to $F_i$.

\subsection{Unified Renderer}
\label{sec:renderer}

\begin{figure*}[t]
\centering
\includegraphics[width=\textwidth]{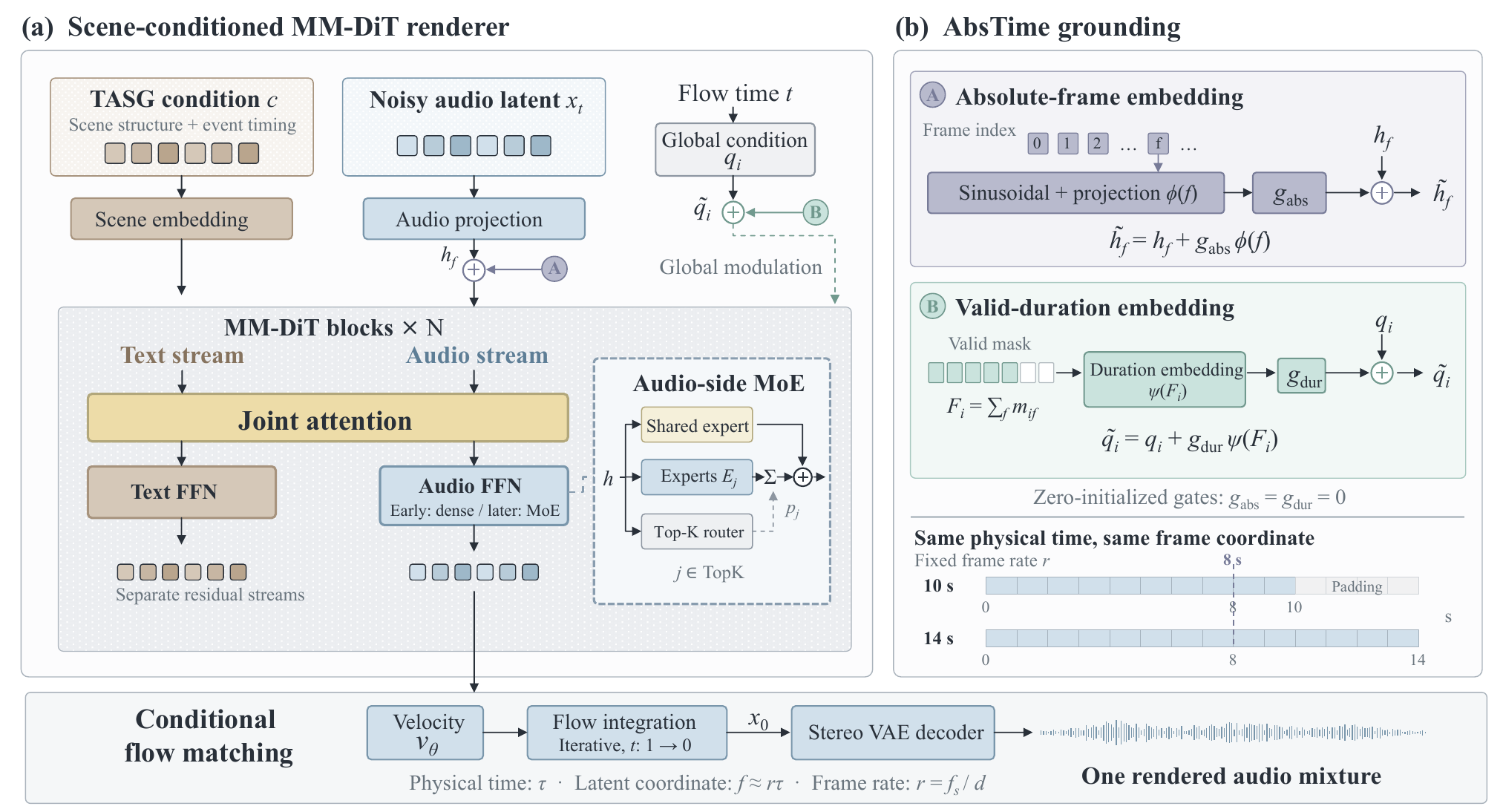}
\caption{SonicRender architecture and AbsTime grounding. (a) A dual-stream MM-DiT combines joint text–audio attention with later audio-side MoE blocks to render one mixture. (b) Zero-gated absolute-frame and valid-duration embeddings provide temporal coordinates: the same physical time retains the same frame index across different canvas lengths.}
\label{fig:architecture}
\end{figure*}

Our renderer is a dual-stream multimodal diffusion transformer (MM-DiT)~\cite{esser2024mmdit}. An audio autoencoder represents the complete mixture as continuous latents $z\in\mathbb{R}^{F\times d_z}$. We use a frozen Qwen3-VL-8B-Thinking model~\cite{qwen3-vl} in text-only mode to encode $c=\mathrm{Serialize}(\mathcal{B})$ into token-level hidden features $U\in\mathbb{R}^{L\times d_u}$, with $d_u=4096$, followed by a learned linear projection to the transformer width. This shared text encoder supplies the conditioning interface across acoustic domains. The audio input projects the noisy latent concatenated with an optional masked reference latent, followed by a residual convolutional position embedding and AbsTime. For text-only generation, the reference latent is zero.

\paragraph{Local interaction and global scene conditioning}
The condition enters at two complementary levels. Joint attention retains token-level access to individual sources and events. In parallel, condition-aware pooling summarizes the scene according to the current flow state and valid duration:
\begin{equation}
\alpha_j = \operatorname{softmax}_{j}\!\left(
\frac{(W_q\tilde q)^\top W_k U_j}{\sqrt{d_u}}\right),\\
q_{\mathrm{scene}} = \tilde q+W_g\sum_{j=1}^{L}\alpha_j U_j .
\label{eq:scene_pool}
\end{equation}
Pooling excludes padded text tokens. The pooled condition modulates both streams through separate AdaLN-Zero modules at every block, as well as the final audio output normalization. A projected $\tilde q$ token is also prepended to the text stream. Thus, detailed event constraints remain individually accessible while a flow- and duration-dependent scene summary conditions the entire rendering computation.

Within each block, text and audio have independent query/key/value and output projections, residual paths, and feed-forward networks. Following the modality-specific coordinate principle of M-RoPE~\cite{qwen2vl}, we use modality-decoupled one-dimensional RoPE, with text and audio positions starting independently. Queries and keys are RMS-normalized, and a single noncausal attention operation covers the concatenated streams:
\begin{equation}
[O_c;O_x]=\operatorname{softmax}\!\left(
\frac{[Q_c;Q_x][K_c;K_x]^\top}{\sqrt{d_h}}+M\right)[V_c;V_x],
\label{eq:joint_attention}
\end{equation}
where subscripts $c$ and $x$ denote the text-condition and audio streams, $d_h$ is the attention head dimension and $M$ masks padding. Outputs are split back into their respective streams. Audio rotary indices are independent of caption length. This block couples source descriptors, event content, and timing with the evolving mixture while preserving modality-specific transformations.

Joint attention provides a computational path for combining the shared and event-specific information separated by \tasg{}. For example, audio frames for a recurring speaker can access the same source description but different event records for each turn's words and delivery. Global modulation supplies scene-wide context, while AbsTime provides frame coordinates for learning the placement of each event. The renderer can thus combine reusable identity information with local variation without relying solely on a pooled scene summary.

Training follows conditional flow matching~\cite{lipman2023flow} on paired scene conditions and target audio mixtures. 
% For data latent $z$, Gaussian noise $\epsilon\sim\mathcal{N}(0,I)$, and flow time $t=\operatorname{sigmoid}(u)$ with $u\sim\mathcal{N}(0,1)$, we form $x_t=(1-t)\epsilon+tz$ and optimize
% \begin{equation}
% \mathcal{L}_{\mathrm{FM}}=\mathbb{E}_{z,\epsilon,t,c}
% \big[\lVert v_\theta(x_t,t,c)-(z-\epsilon)\rVert_{\mathcal{M}}^2\big].
% \label{eq:fm}
% \end{equation}
For clean data latent $x_0 = z$, Gaussian noise $\epsilon \sim \mathcal{N}(0, I)$, and flow time $t = \mathrm{sigmoid}(u)$ with $u \sim \mathcal{N}(0, 1)$, we form $x_t = (1 - t)x_0 + t\epsilon$ and optimize
\begin{equation}
\mathcal{L}_{\mathrm{FM}} = \mathbb{E}_{x_0, \epsilon, t, c}\!\left[\left\| v_{\theta}(x_t, t, c) - (\epsilon - x_0) \right\|_{\mathcal{M}}^{2}\right].
\end{equation}
The squared error is averaged over channels and valid prediction frames $\mathcal{M}$: masked spans when reference audio is retained, or all valid frames when it is dropped. Semantic condition dropout removes both the text-token and pooled-text pathways. Supervision is applied to the complete mixture, letting gradients through joint attention associate source/event requirements with their acoustic realization. Timing is learned within the same generation objective, with AbsTime making positions comparable across scene durations.

\paragraph{Capacity for heterogeneous events}
Complex scenes combine changing speech, transient effects, and slowly varying music or ambience. We use shared-and-routed mixture-of-experts (MoE) feed-forward blocks in later audio layers, retaining dense early layers and joint attention across both streams. For audio hidden state $h$, the routed computation is
\begin{equation}
\mathrm{MoE}(h)=\gamma_s E_{\mathrm{shared}}(h)+
\gamma_r\sum_{j\in\mathrm{TopK}(g(h))}p_j(h)E_j(h),
\label{eq:moe}
\end{equation}
where $g$ is a token-wise linear router and $p_j$ are softmax weights renormalized over the selected experts. The learnable scales start at $(\gamma_s,\gamma_r)=(1,0)$, introducing routed capacity gradually alongside a shared acoustic transformation. Text feed-forward blocks remain dense. 

A standard auxiliary load-balancing loss~\cite{fedus2022switch}, computed only over valid audio tokens and summed across routed layers, supplements flow matching:
\begin{equation}
\mathcal{L}=\mathcal{L}_{\mathrm{FM}}+\lambda_{\mathrm{bal}}\mathcal{L}_{\mathrm{bal}}, \quad
\mathcal{L}_{\mathrm{bal}}=\sum_{l\in\mathcal{L}_{\mathrm{MoE}}} N_e\sum_{j=1}^{N_e}\bar p_j^{(l)}\bar\ell_j^{(l)},
\end{equation}
where $N_e$ is the number of routed experts per layer, $\bar p_j^{(l)}$ is the mean router probability for expert $j$, and $\bar\ell_j^{(l)}$ is the fraction of valid audio tokens routed to expert $j$ (normalized by $K$).

Whereas joint attention integrates scene requirements, audio-side MoE supplies capacity to realize their acoustic combination. A single mixture interval may contain speech, sustained music, and transient effects, requiring heterogeneous transformations within the same latent sequence. Placing MoE in later blocks lets routing use audio states already enriched by source, event, and timing information. The shared expert supplies a common transformation, while routed experts add input-dependent capacity for these local acoustic patterns. Figure~\ref{fig:architecture} illustrates the complete architecture.

% For each routed layer, the balancing term is $N\sum_{j=1}^{N}\bar p_j\bar\ell_j$, where $N$ is the number of routed experts, $\bar p_j$ is the mean router probability, and $\bar\ell_j$ is the mean top-$K$ selection frequency divided by $K$, both averaged over valid audio tokens.

\subsection{Inference}
The model accepts a user-supplied \tasg{} or a scene compiled from natural language, with the requested duration defining the latent canvas. We adopt Adaptive Projected Guidance (APG)~\cite{sadat2025apg} in the clean-latent space induced by CFM. Conditional and text-dropped velocities are converted to clean estimates; their difference is reverse-momentum smoothed and decomposed parallel and orthogonal to the conditional estimate. The parallel component is attenuated, and the resulting update is added to the conditional estimate before conversion back to velocity space. Both branches retain the same AbsTime coordinates and duration. 
% Starting from Gaussian noise at $t=0$, Euler integration proceeds to $t=1$, and the latent is decoded into one complete audio mixture.
% Starting from Gaussian noise $x_1 \sim \mathcal{N}(0, I)$ at $t = 1$, Euler integration proceeds to $t = 0$, and the resulting clean latent $x_0$ is decoded into one complete audio mixture.
Starting from Gaussian noise $x_1 \sim \mathcal{N}(0, I)$ at $t = 1$, Euler integration proceeds to $t = 0$ over 64 steps with a guidance scale of 4.0, and the resulting clean latent $x_0$ is decoded into one complete audio mixture.

\section{Experimental Setup}
\label{sec:exp_setup}

\subsection{Implementation Details}
\label{sec:implementation}

We train \methodname{} on approximately 30,000 hours of eight million internal audio clips for 800,000 optimization steps with a global batch size of 256. Audio is represented by 64-channel latents from a pretrained stereo audio autoencoder at 44.1\,kHz with a downsampling factor of 2,048 (approximately 21.53 latent frames/s). Text features are precomputed by averaging two Qwen hidden layers (indices $-8$ and $-28$) followed by layer normalization; the renderer accepts up to 1,024 text tokens. The MM-DiT has 24 blocks, hidden width 1536, and 24 attention heads. Its last four audio blocks use one shared expert and eight routed experts with top-2 selection. The resulting MoE8 renderer contains approximately 3.0B total parameters with 2.3B active parameters per audio token (matching MoE4 in active compute, compared with 2.2B for the Dense backbone). We use AdamW with a peak learning rate of $3\times10^{-5}$, linear warmup and decay, and $\lambda_{\mathrm{bal}}=0.01$.
% Implementation sources: scripts/train_dual_kling.sh (LR and audio settings),
% infer/infer_dual_moe_kling_mp.py (16 blocks, width 2048, MoE8/top-2),
% te/qwen3vl_te.py (text features), model/trainer.py (optimizer/schedule).
% Before submission, insert the final sampling budget/guidance values from the
% evaluation launch configuration; local launchers contain different settings.

\subsection{Models, Benchmarks, and Protocol}
\label{sec:public_benchmarks}
\label{sec:models}

The full \methodname{} model uses \tasg{}, both AbsTime pathways, and an MoE8 MM-DiT renderer. We evaluate the same checkpoint on Seed-TTS-Eval~\cite{seedttseval2024} (English WER and Mandarin CER), AudioCaps~\cite{kim2019audiocaps} (FAD, KL, and CLAP), MusicCaps~\cite{agostinelli2023musiclm} (KL and CLAP), and MMAG~\cite{zheng2026mmag}. 
To assess multilingual speech generation beyond English and Mandarin, we additionally evaluate the Japanese (JA), Korean (KO), and Spanish (ES) subsets of CV3-Eval~\cite{du2025cosyvoice3}. We use the same checkpoint and text-conditioned synthesis setup as in our Seed-TTS-Eval experiments, and report CER for Japanese and Korean and WER for Spanish (\%), following the benchmark's language-specific scoring units. 
% Published CV3-Eval scores for Fish Audio S2 and VoxCPM2 are included as task-specialist references. \footnote{Baseline scores are taken from the CV3-Eval results in the VoxCPM project repository: \url{https://github.com/OpenBMB/VoxCPM}.}
% CV3-Eval utils/run_wer.py scores JA/KO at character level despite the variable name "wer".
% Confirm that the in-house evaluation retained this official scoring convention.

% MMAG contains 3,974 main-set examples and 1,828 timestamp-subset examples; Table~\ref{tab:mmag} reports the corresponding official main-set KL/WER/CLAP/PC and timestamp-subset WER/CLAP/speech-F1/sound-F1 metrics.
MMAG contains 3,974 main-set examples and 1,828 timestamp-subset examples. Following its official protocol, we report representative metrics: KL, CLAP, WER, and AudioBox Aesthetics production complexity (PC, which reflects multi-source richness), alongside segment-level temporal alignment (Speech-F1 and Sound-F1) on the timestamp subset. Because MMAG specifies event boundaries at sub-second decimal precision (0.01\,s resolution) whereas our canonical \tasg{} representation uses integer-second clock tokens, we round requested boundaries to the nearest integer second during inference and report segment-level F1 under two evaluation references: (i) MMAG's original fractional-second ground-truth windows (primary black scores), which test end-to-end temporal accuracy despite input rounding, and (ii) integer-quantized windows matching the rounded prompts (gray scores), which isolate adherence to the exact conditioned intervals.

Our external comparisons include task specialists and unified audio generators listed in Tables~\ref{tab:single_tasks}--\ref{tab:scenemix}. For mixed-scene experiments, each external model receives a structured rendition of the same scene in its native supported input format, with event content, source identities, relationships, and timing expressed through the controls available to that system. This native-format protocol makes use of the models' available conditioning capabilities. The internal natural-language baseline and \tasg{} model share the MoE8 backbone, AbsTime, training and evaluation scene information, exact utterances, and metric timestamps. The representation comparison therefore changes the organization and binding of the scene instructions while retaining their underlying semantic content. All SceneMix comparisons use reviewed scene specifications, and no generation system accesses evaluation audio or hidden labels. We use three generation seeds (0, 42, and 666) for in-house objective evaluations and seed 42 for human-listening samples.

\subsection{SceneMix: Compositional Multi-Event Scenes}
\label{sec:scenemix}

To evaluate compositional and temporal fidelity, we construct SceneMix, a suite of 168 manually authored multi-event audio scene requests. Each request specifies 2–8 events with temporal windows and acoustic relationships. 150 samples combine speech with non-speech layers spanning music, sound effects, and ambience: 40 feature one speaker, 50 feature two-speaker exchanges, and 60 involve at least three speakers. The remaining 18 contain music and sound effects without speech. Scene durations range from 6 to 12\,s, covering recurring sources, persistent backgrounds, transient effects, overlap, and late returns.

For each generator, a prompt-rendering LLM uses model-specific instructions to convert the authored description into its supported input format: \tasg{} for \methodname{} and the corresponding structured prompts for comparison systems. These renditions preserve the requested content, source relationships, timestamps, and verbatim speech transcripts, enabling comparison of how each system realizes the same multi-event scene.

\subsection{Scene-Fidelity Metrics}
\label{sec:judge}

SceneMix measures event completeness, source binding, and temporal placement. We use Gemini 3.1 Pro Preview under a fixed event-wise protocol to assess each requested event against the generated audio, with separate content and timing judgments. Let $\mathcal{Q}$ denote requested audible events and $d_k=1$ when event $k$ is judged fulfilled, and zero otherwise. We report
\begin{equation}
\mathrm{Event\_Recall}=\frac{\sum_{k\in\mathcal{Q}}d_k}{|\mathcal{Q}|},\
\mathrm{Sound\_Recall}=\frac{\sum_{k\in\mathcal{Q}_{\mathrm{ns}}}d_k}{|\mathcal{Q}_{\mathrm{ns}}|},
\label{eq:recall}
\end{equation}
where $\mathcal{Q}_{\mathrm{ns}}$ comprises non-speech events, including sound effects, music, and ambience. Event Recall measures coverage of the complete composition; Sound Recall separately captures its non-speech layers. WER uses an independent blind transcription pass. Speaker accuracy (SpkAcc) is the fraction of requested speech-turn pairs whose observed same/different-speaker relation agrees with the source IDs, with unmatched pairs counted as failures.

For timed targets $\mathcal{Q}_t$, $w_k=1$ when the event is judged aligned with its requested window. The judgment follows the event's temporal extent: a brief event should occur within the permitted region, while a continuous background should persist across its requested span. This measures fulfillment of the temporal instruction while accommodating the different durations of speech, effects, and background layers. We report
\begin{equation}
\mathrm{WHR}=\frac{\sum_{k\in\mathcal{Q}_t} d_k w_k}{|\mathcal{Q}_t|},
\mathrm{CWA}=\frac{\sum_{k\in\mathcal{Q}_t} d_k w_k}{\sum_{k\in\mathcal{Q}_t}d_k}.
\label{eq:joint_whr}
\end{equation}
WHR scores content and timing jointly over all timed targets; CWA scores timing among fulfilled timed targets. Partial or uncertain content judgments remain unsuccessful targets. Their relationship is $\mathrm{WHR}=\mathrm{CWA}\times\mathrm{Recall}(\mathcal{Q}_t)$ for the same pooled counts. All models use the same evaluation protocol. The separate speech-onset analysis extracts word-level onset timestamps $\hat b_k$ via WhisperX forced alignment and tests $|\hat b_k-b_k|\leq\delta$, counting missing onsets as failures; together with MMAG segment F1, it complements window compliance with finer temporal localization and interval agreement.

In addition, Gemini 3.1 Pro Preview rates overall technical quality (AI-Tech) and whole-scene instruction adherence (AI-Sem) on a 1--5 scale.

\subsection{Human Evaluation}
\begin{table*}[t]
\centering
\small
\caption{Public single-task generation. Seed-TTS-Eval reports English WER and Mandarin CER (\%); AudioCaps uses FAD, KL, and CLAP; MusicCaps uses KL and CLAP. Dashes indicate unavailable entries.}
\label{tab:single_tasks}
\begin{tabular}{lccccccc}
\toprule
& \multicolumn{2}{c}{Seed-TTS-Eval (TTS)}
& \multicolumn{3}{c}{AudioCaps (TTA)}
& \multicolumn{2}{c}{MusicCaps (TTM)} \\
\cmidrule(lr){2-3}\cmidrule(lr){4-6}\cmidrule(lr){7-8}
Model & EN WER$\downarrow$ & ZH CER$\downarrow$
& FAD$\downarrow$ & KL$\downarrow$ & CLAP$\uparrow$
& KL$\downarrow$ & CLAP$\uparrow$ \\
\midrule
\multicolumn{8}{l}{\emph{Task-specific generators}} \\
F5-TTS~\cite{chen2024f5tts}
& 1.8\% & 1.6\% & -- & -- & -- & -- & -- \\
CosyVoice 3~\cite{du2025cosyvoice3}
& 1.5\% & 0.7\% & -- & -- & -- & -- & -- \\
Qwen3-TTS~\cite{qwen2026tts}
& 1.2\% & 0.8\% & -- & -- & -- & -- & -- \\
Higgs Audio V2~\cite{boson2025higgsv2}
& 1.0\% & 0.8\% & -- & -- & -- & -- & -- \\
TangoFlux~\cite{hung2026tangoflux}
& -- & -- & \textbf{2.26} & \textbf{1.15} & 0.480 & -- & -- \\
MusicGen-Large~\cite{copet2023musicgen}
& -- & -- & -- & -- & -- & 1.31 & 0.280 \\
AudioX~\cite{tian2026audiox}
& -- & -- & 2.45 & 1.27 & 0.440 & \textbf{0.96} & 0.240 \\
AudioLDM 2~\cite{liu2024audioldm2}
& -- & -- & 2.29 & 1.41 & 0.419 & 1.20 & 0.301 \\
Stable Audio Open~\cite{evans2024stableopen}
& -- & -- & 4.13 & 2.14 & 0.350 & 1.44 & 0.300 \\
UniFlow-Audio~\cite{xu2025uniflow}
& -- & -- & 5.74 & 1.43 & 0.476 & 1.87 & 0.241 \\
\midrule
\multicolumn{8}{l}{\emph{Unified generators}} \\
Ming-Omni-TTS-0.5B~\cite{ming2026omnittts}
& 2.2\% & 0.9\% & -- & 2.26 & 0.347 & -- & -- \\
Ming-Omni-TTS-16.8B~\cite{ming2026omnittts}
& 2.0\% & 0.8\% & -- & 1.64 & 0.424 & -- & -- \\
Dasheng AudioGen~\cite{mei2026dasheng}
& 12.2\% & -- & 3.19 & 1.86 & 0.438 & 1.37 & 0.334 \\
UniMoE-Audio~\cite{liu2026unimoe}
& 1.3\% & 0.8\% & -- & -- & -- & 1.25 & 0.340 \\
UniAudio 2.0~\cite{yang2026uniaudio2}
& 3.6\% & 2.3\% & -- & 3.26 & 0.170 & 1.80 & -- \\
MiDashengLM-Gen~\cite{sun2026midasheng}
& 2.8\% & 3.9\% & 5.01 & 1.91 & 0.360 & 1.24 & 0.330 \\
SonicWeave~\cite{cai2026sonicweave}
& 1.0\% & 0.8\% & 2.75 & 1.26 & 0.475 & 1.06 & 0.312 \\
\midrule
\multicolumn{8}{l}{\emph{Our models}} \\
Natural language + AbsTime
& 3.9\% & 3.2\% & 3.17 & 1.56 & 0.432 & 0.99 & 0.310 \\
\methodname{}
& \textbf{0.9\%} & \textbf{0.6\%} & 2.58 & \textbf{1.15} & \textbf{0.507} & 0.97 & \textbf{0.384} \\
\bottomrule
\end{tabular}
\end{table*}
\label{sec:human_eval}
% We select 20 scenes from SceneMix for subjective evaluation of \methodname{} and five comparison systems, yielding 120 generated clips. 
We select a representative subset of 20 scenes from SceneMix for subjective evaluation of \methodname{} and five comparison systems, which spans one- to multi-speaker dialogues, non-speech compositions, and event loads from 2 to 8 concurrent or sequential activities.
Fifteen listeners rate anonymous, randomized samples against the same natural-language scene descriptions on two five-point scales. MOS-Q measures overall audio quality, including clarity, naturalness, and freedom from artifacts; MOS-R measures scene realization and relevance, including requested content, events, source relationships, and foreground/background composition. Higher scores are better. Temporal order and overlap contribute to MOS-R when they affect scene meaning, while precise timestamp agreement is assessed separately by the temporal metrics. Listening samples use seed 42. We report mean scores with 95\% confidence intervals and conduct pairwise significance tests (Wilcoxon signed-rank test, $p < 0.01$) between \methodname{} and each comparison system on both dimensions.

\section{Results and Analysis}

\label{sec:results}

\subsection{Unified Generation Across Speech, Sound, and Music}
\label{sec:single_task_results}

% Table~\ref{tab:single_tasks} tests whether improved scene-level control comes at the expense of the single-task capabilities that make a generator genuinely unified. 
% Table~I demonstrates that SonicRender not only excels at complex scene rendering, but also outperforms both unified baselines and several task-specific specialists across standard speech, sound, and music benchmarks using a single checkpoint.
Table~I demonstrates that SonicRender outperforms both unified baselines and several task-specific specialists across standard speech, sound, and music benchmarks using a single checkpoint.
\methodname{} achieves 0.9\% English WER and 0.6\% Mandarin CER on Seed-TTS-Eval, the lowest values among the unified generators listed in the table. The matched natural-language model reaches 3.9\%/3.2\%, showing that the structured scene condition improves lexical realization even for the relatively simple one-task setting. Speaker/source binding is therefore not only useful in complex mixtures: explicit content organization also provides a stronger conditioning interface for ordinary speech synthesis.

On AudioCaps, \methodname{} obtains 0.507 CLAP and 1.15 KL, compared with 0.432 and 1.56 for its natural-language counterpart. 
On MusicCaps, \methodname{} achieves the highest CLAP (0.384) across the entire table and the lowest KL (0.97) among all unified generators (closely matching the specialist AudioX at 0.96).
Rather than trading off single-domain quality for multi-event control, SonicRender establishes superior distributional fidelity and text--audio alignment across all three acoustic regimes.
% On MusicCaps, it reaches 0.384 CLAP and 0.97 KL. The AudioCaps CLAP score is the highest listed, while the MusicCaps CLAP score likewise exceeds the other reported entries. These comparisons are informative because they cover distinct acoustic regimes: intelligible lexical sequences, environmental sound texture, and extended musical structure. The results show that the scene-oriented renderer retains competitive distributional measures while strengthening text--audio correspondence across all three types. Task-specialist speech results may use reference prompting and are included as task-level context, as specified in the evaluation protocol.

\begin{table}[h]
\centering
\small
\caption{Multilingual speech generation on CV3-Eval. }
\label{tab:multilingual}
\setlength{\tabcolsep}{7pt}
\begin{tabular}{lccc}
\toprule
Model & JA CER$\downarrow$ & KO CER$\downarrow$ & ES WER$\downarrow$ \\
\midrule
Fish Audio S2~\cite{liao2026fishaudios2} & \textbf{3.96\%} & 2.76\% & \textbf{2.00\%} \\
VoxCPM2~\cite{zhou2026voxcpm2} & 5.96\% & 5.69\% & 3.80\% \\
\midrule
\methodname{} & 5.23\% & \textbf{2.27\%} & 4.78\% \\
\bottomrule
\end{tabular}
\end{table}

Table~\ref{tab:multilingual} extends the speech evaluation beyond English and Mandarin. Using the same unified checkpoint, \methodname{} achieves CERs of 5.23\% and 2.27\% on the Japanese and Korean subsets of CV3-Eval, respectively, and a WER of 4.78\% on Spanish. These low transcription error rates demonstrate accurate speech content realization across the three languages, extending the linguistic coverage of the unified renderer. Our demo page further showcases additional languages, Chinese dialects and regional varieties (e.g., Cantonese, Sichuanese, and Northeastern Mandarin), and English accents (e.g., British and Indian English), including examples with accompanying music, ambience, and sound effects. 
% We invite readers to explore these examples at \url{https://caiyunrui.github.io/SonicRender}.

\subsection{Timestamped Mixed-Audio Generation on MMAG}
\begin{table*}[t]
\centering
\small
\caption{MMAG main-set and timestamp-subset generation. During inference, our models round requested timestamps to integer seconds; segment-level F1 is evaluated against both MMAG's original fractional-second ground-truth windows (primary black scores) and the corresponding integer-quantized windows (gray scores).}
\label{tab:mmag}
\begin{tabular}{lcccccccc}
\toprule
& \multicolumn{4}{c}{Main: overall captions} & \multicolumn{4}{c}{Timestamp: timed captions} \\
\cmidrule(lr){2-5}\cmidrule(lr){6-9}
Model & KL$\downarrow$ & WER$\downarrow$ & CLAP$\uparrow$ & PC$\uparrow$ & WER$\downarrow$ & CLAP$\uparrow$ & Speech-F1$\uparrow$ & Sound-F1$\uparrow$ \\
\midrule
AuDirector (agentic)~\cite{ren2026audirector} & 1.60 & 0.08 & 0.34 & 3.69 & 0.15 & 0.35 & 0.72 & 0.57 \\
LTX-2 (audio-visual)~\cite{hacohen2026ltx2} & 1.31 & 0.04 & 0.34 & 3.04 & 0.05 & 0.33 & 0.64 & 0.44 \\
UniAVGen (audio-visual)~\cite{zhang2026uniavgen} & 2.28 & 0.10 & 0.25 & 2.35 & 0.15 & 0.23 & 0.63 & 0.29 \\
\midrule
Dasheng AudioGen-Base~\cite{mei2026dasheng} & \textbf{0.87} & 0.11 & 0.34 & 3.55 & 0.43 & 0.31 & 0.60 & 0.41 \\
Dasheng AudioGen-Fine~\cite{mei2026dasheng} & 0.89 & 0.11 & 0.33 & 3.64 & 0.13 & 0.31 & 0.62 & 0.42 \\
Ming-Omni-TTS-0.5B~\cite{ming2026omnittts} & 2.80 & 0.03 & 0.26 & 1.60 & -- & -- & -- & -- \\
Ming-Omni-TTS-16.8B~\cite{ming2026omnittts} & 2.46 & 0.03 & 0.27 & 1.79 & -- & -- & -- & -- \\
\midrule
Natural language + AbsTime & 0.98 & 0.07 & 0.32 & 4.05 & 0.08 & 0.28 & 0.72 / \textcolor{gray}{0.75} & 0.50 / \textcolor{gray}{0.54} \\
\methodname{} & 0.90 & \textbf{0.02} & \textbf{0.40} & \textbf{4.21} & \textbf{0.03} & \textbf{0.39} & \textbf{0.79} / \textcolor{gray}{\textbf{0.86}} & \textbf{0.60} / \textcolor{gray}{\textbf{0.65}} \\
\bottomrule
\end{tabular}
\end{table*}
\label{sec:mmag_results}

\begin{table*}[t]
\centering
\caption{SceneMix evaluation: event and non-speech completeness, content, timing, speaker binding, and perceptual/semantic quality. WHR scores every requested timed event, counting omissions as failures. MOS is evaluated on the 20-scene listening subset and reported as mean $\pm$ 95\% confidence-interval half-width.}
\label{tab:scenemix}
\small
\setlength{\tabcolsep}{4.2pt}
\begin{tabular}{lccccccccc}
\toprule
& \multicolumn{5}{c}{Objective} & \multicolumn{2}{c}{AI judge} & \multicolumn{2}{c}{Human MOS} \\
\cmidrule(lr){2-6}\cmidrule(lr){7-8}\cmidrule(lr){9-10}
Model & Sound Rec.$\uparrow$ & Event Rec.$\uparrow$ & WER$\downarrow$ & WHR$\uparrow$ & SpkAcc$\uparrow$ & AI-Tech$\uparrow$ & AI-Sem$\uparrow$ & MOS-Q$\uparrow$ & MOS-R$\uparrow$ \\
\midrule
Dasheng AudioGen-Fine~\cite{mei2026dasheng} & 56.4\% & 59.5\% & 10.2\% & 40.8\% & 49.0\% & 3.10 & 2.95 & 2.74$_{\pm 0.32}$ & 2.88$_{\pm 0.30}$ \\
MiDashengLM-Gen~\cite{sun2026midasheng} & 62.6\% & 75.5\% & 8.1\% & 48.7\% & 70.8\% & 3.27 & 3.20 & 3.52$_{\pm 0.35}$ & 3.69$_{\pm 0.34}$ \\
Ming-Omni-TTS-0.5B~\cite{ming2026omnittts} & 37.0\% & 50.2\% & 3.2\% & 35.9\% & 58.7\% & 4.14 & 2.84 & 3.55$_{\pm 0.32}$ & 3.28$_{\pm 0.41}$ \\
Ming-Omni-TTS-16.8B~\cite{ming2026omnittts} & 48.4\% & 53.0\% & 2.7\% & 34.6\% & 65.8\% & 4.32 & 2.96 & 3.82$_{\pm 0.27}$ & 3.37$_{\pm 0.30}$ \\
\midrule
Natural language + AbsTime & 59.9\% & 74.6\% & 5.2\% & 68.0\% & 86.6\% & 4.29 & 3.93 & 3.90$_{\pm 0.23}$ & 3.74$_{\pm 0.34}$ \\
\methodname{} & \textbf{71.1\%} & \textbf{85.4\%} & \textbf{1.1\%} & \textbf{83.9\%} & \textbf{95.7\%} & \textbf{4.53} & \textbf{4.34} & \textbf{4.25}$_{\pm 0.20}$ & \textbf{4.44}$_{\pm 0.25}$ \\
\bottomrule
\end{tabular}
\end{table*}

MMAG provides a public mixed-scene test that simultaneously exposes content quality and timestamp following (Table~\ref{tab:mmag}). On the main set, \methodname{} reaches 0.02 WER, 0.40 CLAP, and 4.21 PC, improving on the matched natural-language model's 0.07 WER, 0.32 CLAP, and 4.05 PC. The strong CLAP and PC scores are particularly relevant in mixtures, where a fluent speech track alone may leave music or transient requirements unsatisfied. 
The overall KL score is 0.90, close to Dasheng AudioGen-Base's 0.87, indicating that the gains in instruction adherence are accompanied by competitive distributional matching.

The timestamp subset makes the time-control claim explicit. Because our training corpus standardizes \tasg{} clock tokens at integer-second resolution, we round MMAG's decimal timestamps to the nearest integer second during inference to match the training token distribution. Remarkably, even with the inherent boundary error ($\le 0.5$\,s), evaluating the generated mixtures directly against the benchmark's original fractional-second ground-truth windows yields 0.79 speech-F1 and 0.60 sound-F1, surpassing all compared systems. Its timestamped-set CLAP also reaches 0.39, outperforming the matched natural-language model's 0.28. This confirms that \methodname{} generates recognizable heterogeneous events and places them accurately on the timeline despite prompt-level quantization.

Evaluating the same generated outputs against integer-quantized ground-truth windows (gray entries) further isolates the renderer's intrinsic adherence to its conditioned schedule. Under this resolution-matched evaluation, speech and sound F1 rise from 0.79/0.60 to 0.86/0.65 (and from 0.72/0.50 to 0.75/0.54 for the natural-language variant). 
Since AbsTime itself operates at a fine latent frame rate of 21.53\,Hz ($\approx 46$\,ms per frame), the gap between the black and gray scores reflects only the text-side token convention rather than an architectural limit, while the leading performance under both criteria confirms precise metric-time grounding.

% The two numerical resolutions also reveal an important property of the learned time representation. Our canonical TASG input format uses integer-second time tokens; gray entries report inference after converting the MMAG instructions to this format. Under the integer-second rendition, speech/sound F1 reaches 0.86/0.65, compared with 0.79/0.60 for the original fractional-second rendition. The principal result is the latter: the model already leads the reported baselines when it is given the benchmark's native decimal timestamps. Strong performance at both resolutions supports the ability to associate symbolic time values with identifiable acoustic activity, while the additional increase under integer quantization is consistent with the granularity used by the canonical condition.

\subsection{Scene Fidelity on SceneMix}
\label{sec:scene_results}

SceneMix measures whether a request is rendered as a complete scene, including its less dominant acoustic components. In Table~\ref{tab:scenemix}, \methodname{} achieves 85.4\% all-event recall and 71.1\% Sound Recall, together with 1.1\% English WER and 95.7\% speaker accuracy. The gap between the two recall scores is meaningful: non-speech components such as effects, music, and ambience are more vulnerable to omission in dialogue-centered scenes, so reporting only all-event recall would conceal a major aspect of scene fidelity. Among the external systems, MiDashengLM-Gen reaches 75.5\% all-event recall and 62.6\% Sound Recall; \methodname{} improves both by 9.9 and 8.5 percentage points, respectively.

Temporal and identity metrics reinforce the same picture. SonicRender achieves 83.9\% joint window-hit rate (WHR), substantially higher than the 48.7\% of MiDashengLM-Gen, the strongest external entry in that column. This means that the measured event is not merely audible: it is also placed within its permitted time window. Speaker accuracy is 95.7\%, compared with 70.8\% for MiDashengLM-Gen, reflecting the challenge of retaining source identities across several local realizations. 
% AI-Tech of 4.53 and AI-Sem of 4.34 indicate that this stronger constraint fulfillment also coincides with favorable holistic assessment. 
SonicRender also achieves the highest AI-Tech (4.53) and AI-Sem (4.34) scores, confirming that strict event and timing control directly enhances overall perceptual quality and semantic faithfulness.

\begin{table*}[t]
\centering
\caption{Ablations of structured conditioning, AbsTime, and audio-side capacity. CWA scores window placement among detected events; WHR scores all timed targets. In each MMAG F1 cell, black/gray scores denote evaluation against original fractional-second and integer-quantized ground-truth windows, respectively.}
\label{tab:core_ablation}
\small
\setlength{\tabcolsep}{4.5pt}
\begin{tabular}{lccccccccc}
\toprule
& & \multicolumn{4}{c}{SceneMix: content and binding} & \multicolumn{2}{c}{Temporal} & \multicolumn{2}{c}{MMAG timestamp} \\
\cmidrule(lr){3-6}\cmidrule(lr){7-8}\cmidrule(lr){9-10}
Configuration & Backbone & Sound Rec.$\uparrow$ & WER$\downarrow$ & SpkAcc$\uparrow$ & AI-Sem$\uparrow$ & WHR$\uparrow$ & CWA$\uparrow$ & Speech-F1$\uparrow$ & Sound-F1$\uparrow$ \\
\midrule
TASG (w/o AbsTime) & MoE8 & 69.2\% & 1.7\% & 94.8\% & 4.25 & 71.6\% & 85.0\% & 0.70 / \textcolor{gray}{0.73} & 0.48 / \textcolor{gray}{0.53}  \\
Natural language + AbsTime & MoE8 & 59.9\% & 5.2\% & 86.6\% & 3.93 & 68.0\% & 91.2\% & 0.72 / \textcolor{gray}{0.75} & 0.50 / \textcolor{gray}{0.54} \\
TASG + AbsTime (full) & MoE8 & \textbf{71.1\%} & \textbf{1.1\%} & \textbf{95.7\%} & \textbf{4.34} & \textbf{83.9\%} & \textbf{98.3\%} & \textbf{0.79} / \textcolor{gray}{\textbf{0.86}} & \textbf{0.60} / \textcolor{gray}{\textbf{0.65}} \\
\midrule
TASG + AbsTime & MoE4 & 69.0\% & 2.2\% & 94.3\% & 4.19 & 78.9\% & 96.5\% & 0.76 / \textcolor{gray}{0.84} & 0.58 / \textcolor{gray}{0.64}  \\
TASG + AbsTime & Dense & 60.6\% & 4.5\% & 90.6\% & 4.02 & 77.9\% & 96.5\% & 0.75 / \textcolor{gray}{0.84} & 0.56 / \textcolor{gray}{0.61}  \\
\bottomrule
\end{tabular}
\end{table*}

The matched representation comparison is especially informative because both variants receive the same scene facts and use the same MoE8 backbone and AbsTime. Organizing those facts as \tasg{} increases all-event recall by 10.8 percentage points (74.6\% to 85.4\%) and Sound Recall by 11.2 points (59.9\% to 71.1\%). Speaker accuracy rises by 9.1 points, from 86.6\% to 95.7\%, while WHR rises by 15.9 points, from 68.0\% to 83.9\%. English WER falls from 5.2\% to 1.1\%. The improvements span inclusion, lexical content, identity, and temporal placement rather than concentrating in a single easy-to-optimize score. 
They show that explicit source--event assignments and local time windows provide a more usable conditioning structure for a unified renderer.

% \paragraph{Subjective quality and scene realization}
The listening study corroborates these objective gains. \methodname{} achieves MOS-Q of $4.25\pm0.20$ and MOS-R of $4.44\pm0.25$ (95\% confidence intervals), the highest scores among all six systems. Compared with the natural-language + AbsTime variant, the mean gains are 0.35 and 0.70 points, respectively. Pairwise significance tests show that \methodname{} significantly outperforms each of the five comparison systems on both MOS-Q and MOS-R. These results establish that its compositional gains translate into perceptibly better audio quality and more faithful scene realization.

To assess the reliability of Gemini 3.1 Pro Preview as an automated judge, we measure its alignment with human evaluations on the 20-scene listening subset (120 samples across six systems). At the system level, AI-Tech and AI-Sem both achieve a Spearman correlation of $\rho = \text{0.94}$ with MOS-Q and MOS-R, respectively. On a 30-sample stratified audit independently annotated by two human raters, Gemini achieves 91.0\% agreement ($\kappa = \text{0.70}$) on event presence, 88.2\% ($\kappa = \text{0.63}$) on window alignment, and 95.1\% ($\kappa = \text{0.74}$) on speaker-pair consistency, measured against consensus human annotations.

\subsection{Ablations: Structure, Time Grounding, and Capacity}
\label{sec:ablation}

Table~\ref{tab:core_ablation} compares five trained configurations: the full model, its information-matched natural-language variant, a model without AbsTime, and MoE4/Dense alternatives. This matrix separates the contribution of organized scene conditions, the audio-side time reference, and routed capacity at the backbone level. Because the comparisons share the same scored scene requirements, changes in event coverage, content accuracy, and placement can be examined together.

\paragraph{Source-aware scene organization}
With MoE8 and AbsTime fixed, the full structured model surpasses the natural-language condition by 11.2 points in Sound Recall, 9.1 points in speaker accuracy, and 15.9 points in WHR. The gain in all-event recall reported in Table~\ref{tab:scenemix} is also 10.8 points. The joint improvement is important: greater speech intelligibility alone cannot account for better non-speech retention, and an unchanged timing mechanism cannot explain the full difference in source binding. The organization of information---separate records for persistent identities, local activities, and their time windows---is therefore a substantial component of the model's scene-control capability.

\paragraph{Audio-side temporal grounding}
Removing AbsTime lowers WHR from 83.9\% to 71.6\% (12.3 points) and conditional window accuracy from 98.3\% to 85.0\% (13.3 points), although Sound Recall changes by only 1.9 points (71.1\% versus 69.2\%). The close Sound Recall indicates that the remaining representation and backbone still produce many of the requested non-speech events. The larger temporal losses identify where audio-side grounding matters: placing an audible event at the intended moment. On the independent MMAG timed set, speech/sound F1 decreases from 0.79/0.60 to 0.70/0.48 when AbsTime is removed. Both evaluations therefore point to a consistent placement advantage for the explicit audio timeline.

\paragraph{Backbone capacity}
The full MoE8 renderer reaches 71.1\% Sound Recall and 83.9\% WHR, compared with 69.0\%/78.9\% for MoE4 and 60.6\%/77.9\% for Dense. The clearest capacity-dependent change is Sound Recall: an increase of 10.5 points from Dense to MoE8, accompanied by an English WER reduction from 4.5\% to 1.1\% and speaker-accuracy growth from 90.6\% to 95.7\%. MMAG F1 also increases from 0.75/0.56 to 0.79/0.60 when evaluated against the original fractional-second windows.
These results are consistent with routed audio capacity helping represent heterogeneous local acoustic events while the scene structure and AbsTime supply the shared controls.

\subsection{Robustness to Increasing Event Load}
\label{sec:event_load}

\begin{figure}[!tbp]
\centering
\includegraphics[width=\columnwidth]{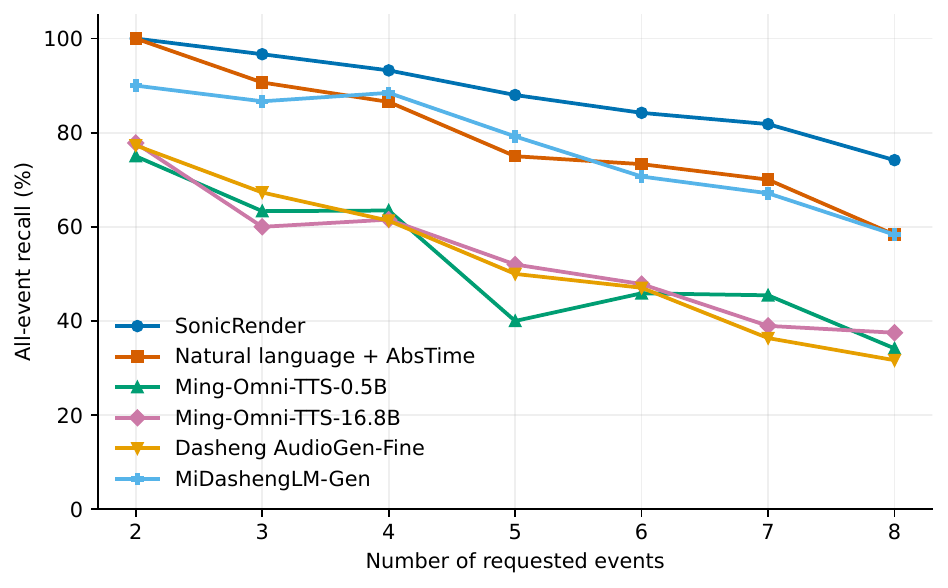}
\caption{Event completeness versus compositional load. Recall is computed over requested speech and non-speech events within each event-count group, with each event contributing equally. Natural language + AbsTime uses the matched backbone and scene information.}
\label{fig:event_load}
\end{figure}

The event-count analysis tests a central prediction of structured scene rendering: as a request contains more independently specified events, source/event binding should help prevent omissions. Figure~\ref{fig:event_load} groups requests by the number of requested events and evaluates the presence of \emph{all} targets within each group. It therefore tracks the completeness of a generated composition, not the correctness of its onset windows.

% At two events, SonicRender and the natural-language + AbsTime baseline both reach 100\% recall, illustrating that relatively simple requests can often be satisfied without a sophisticated organization of the condition. The separation increases with load. 
While simple two-event scenes saturate at 100\% recall for both internal models, the advantage of structured rendering widens sharply as compositional density grows.
At six events, SonicRender retains approximately 84\% of requested events; at eight, its recall remains about 74\%, compared with roughly 58\% for both the matched natural-language baseline and MiDashengLM-Gen in that bin. The curve also shows a steadier decline than competing systems, whose recall can fall sharply once multiple classes of activity must coexist. This behavior is especially relevant for scenes with long-lasting background layers and short inserted events: the number of acoustic components grows, but each remains a separate requested obligation. The measured advantage is therefore not only in average content coverage; it persists as the number of requested events rises.

\subsection{Onset Alignment and Error Tolerance}
\label{sec:onset_analysis}

\begin{figure}[!tbp]
\centering
\includegraphics[width=\columnwidth]{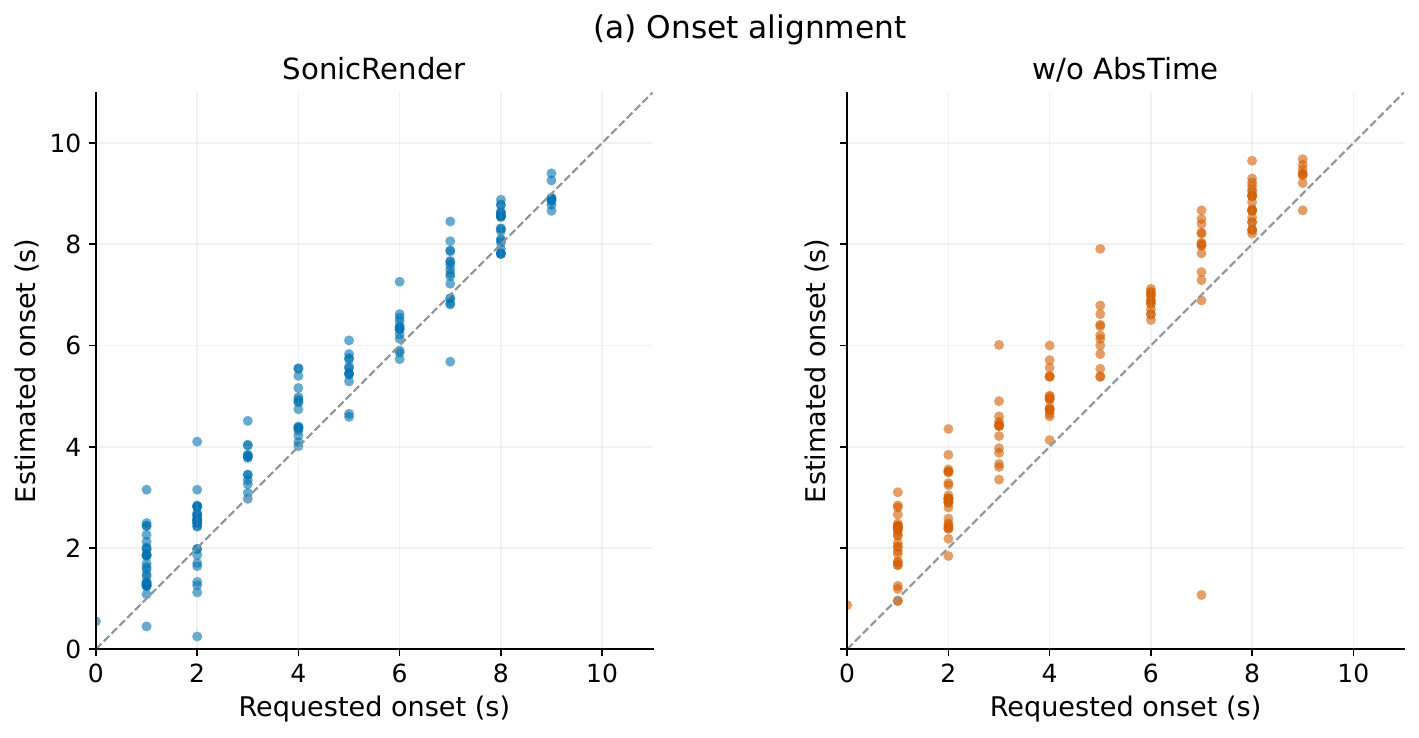}
\vspace{0.4em}
\includegraphics[width=0.8\columnwidth]{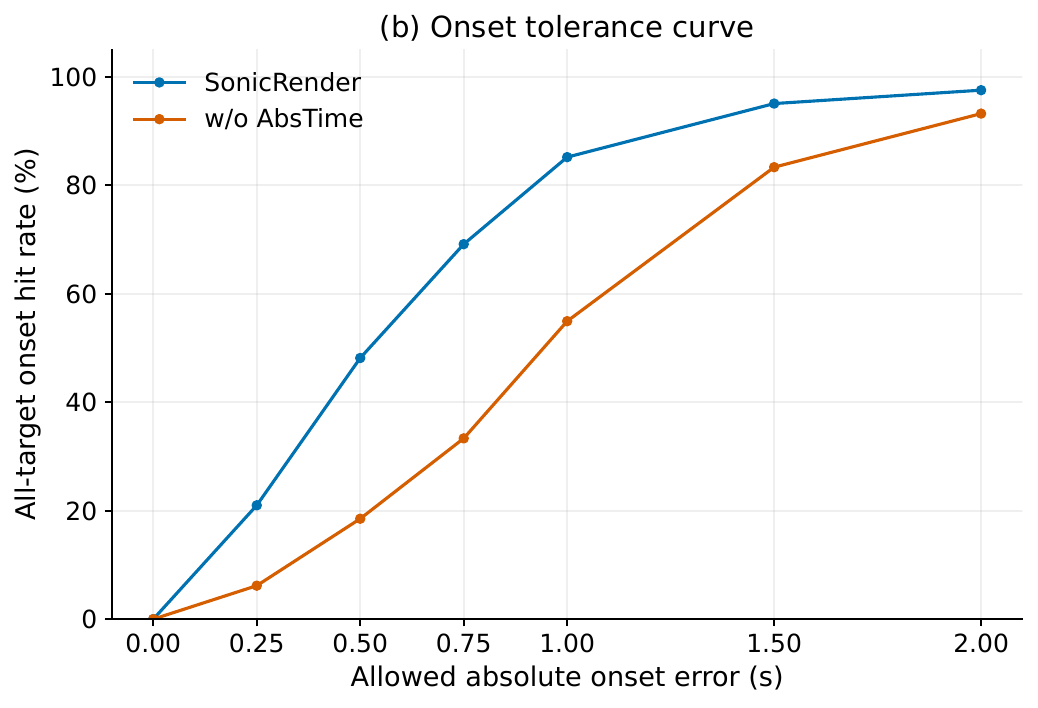}
\caption{Effect of AbsTime on speech-onset alignment under matched TASG inputs. (a) Requested versus estimated onsets for located speech events; dashed diagonals indicate exact alignment. (b) The fraction of all target speech events localized within each absolute-error tolerance. }
\label{fig:onset_analysis}
\end{figure}

The preceding WHR analysis concerns whether timed events are present and satisfy their requested windows. We additionally examine the more localized question of speech onset error under TASG conditioning. Figure~\ref{fig:onset_analysis}(a) plots requested versus automatically estimated onsets, with the dashed diagonal indicating ideal alignment. SonicRender's observations cluster near the diagonal, while the configuration without AbsTime shows a recurrent positive offset, particularly for speech requested later in the scene. The difference is visible across several requested positions, making the effect more informative than a single mean absolute error.

Figure~\ref{fig:onset_analysis}(b) quantifies how often these estimates satisfy increasingly permissive onset-error thresholds. At a 0.5-s tolerance, the hit rate is approximately 48\% for the full model versus 19\% without AbsTime, a difference of 29 points. At 1.0\,s the corresponding rates are 85\% and 55\%, and at 1.5\,s they are 95\% and 83\%. The largest separation occurs at tolerances that require accurate placement: the explicit metric coordinate is most valuable when an event cannot simply occur somewhere in the scene. All target onsets remain in the denominator, so missed targets do not create an artificially high conditional hit rate. These automatically estimated onsets provide a complementary diagnostic to the separately defined SceneMix WHR and CWA.

Together, the scatter and tolerance analyses help explain the grounding ablation in Table~\ref{tab:core_ablation}. Without AbsTime, the model still generates many events, yet tends to place speech too late. With AbsTime, the same symbolic schedule is associated more consistently with audio latent positions. The time-coordinate mechanism therefore improves \emph{where} an event occurs in addition to whether it is generated.

\subsection{Qualitative Analysis of Layered Scenes}
\label{sec:case_study}

\begin{figure}[!tbp]
\centering
\includegraphics[width=\columnwidth]{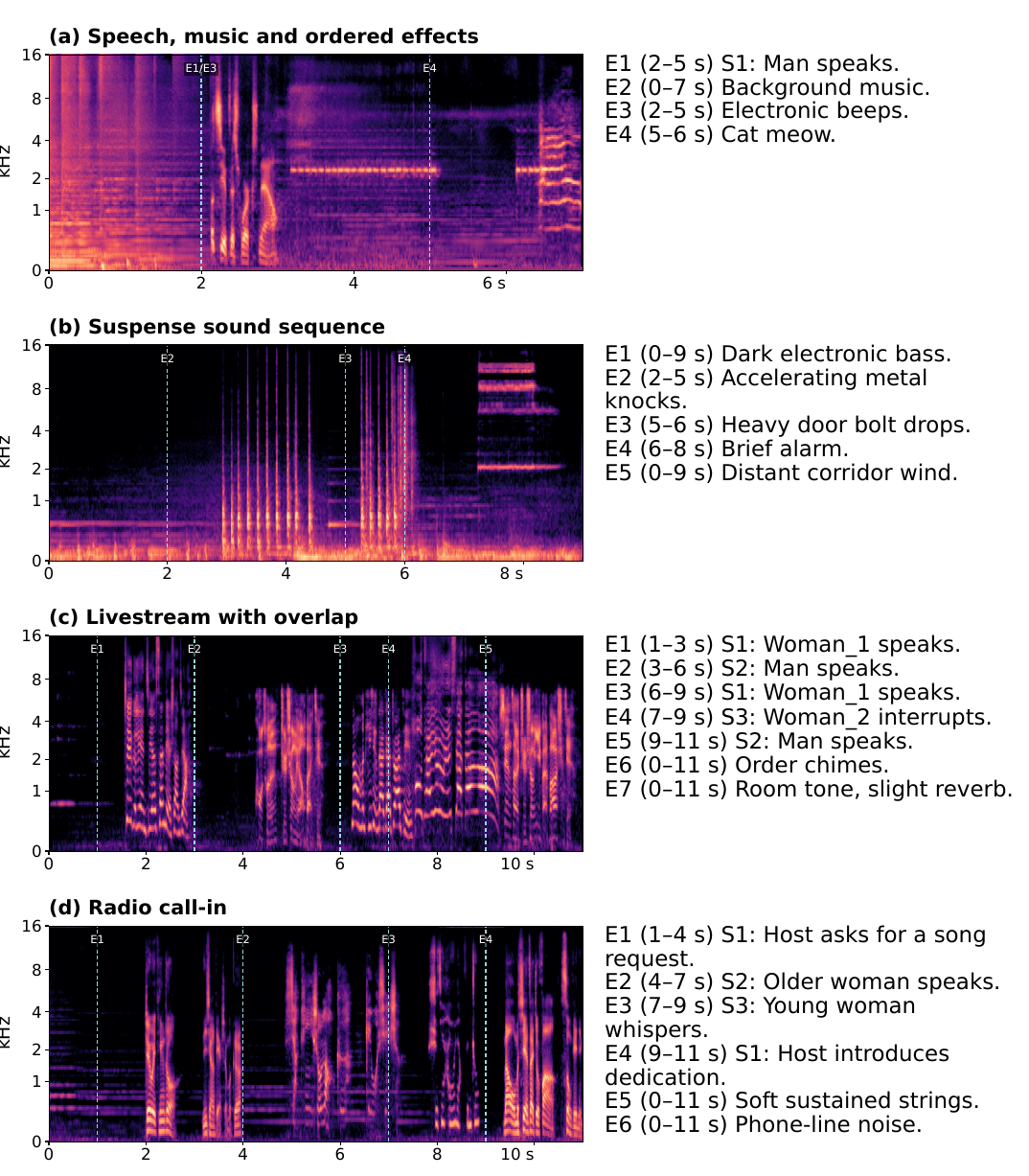}
\caption{Generated multi-event scenes: (a) speech, music, and ordered effects; (b) a suspense sound sequence; (c) overlapping livestream dialogue; and (d) a radio call-in. Mel spectrograms are paired with abridged input specifications. Event labels and dashed vertical lines identify requested onsets; repeated source identifiers indicate recurring speakers.}
\label{fig:cases}
\end{figure}

Panels (a) and (b) of Fig.~\ref{fig:cases} present two scenes that illustrate different forms of acoustic composition. In panel (a), a short utterance, a continuing music layer, repeated electronic beeps, and a brief cat meow have different durations and roles. The sustained music and localized high-energy events are distinguishable in the spectrogram, illustrating the coexistence of extended texture and time-specific sound effects. In panel (b), metallic knocks accelerate before a heavy bolt event and a short alarm, while a low-frequency bass bed and distant wind continue through the sequence. The ordered transient pattern is an example of detailed event structure beyond a global description such as \emph{suspenseful ambience}.

The two conversation cases emphasize source identity across dialogue turns and competing layers. In Fig.~\ref{fig:cases}(c), a woman speaks twice, a man gives stock updates, another woman interjects, and repeated notification chimes remain part of the soundscape. The visible voiced regions and transients occupy different parts of the mixture, including overlapping requirements. Figure~\ref{fig:cases}(d) depicts a radio call-in, where the host returns after two callers against sustained strings and telephone noise. Stable source identifiers bind repeated speech activities to the same requested speaker despite changing lexical content and local delivery.

These examples complement the quantitative evidence by making the acoustic forms involved in complex multi-event generation inspectable: harmonic backgrounds, transient attacks, recurring dialogue, and sustained noisy layers. While spectrograms visualize spectral energy and timing, the recurrence fidelity is measured separately by SpkAcc and the listening protocol. Together, the four scenes illustrate the heterogeneous, temporally organized content addressed by the event-completeness and temporal-grounding evaluations.

\section{Conclusion}
\label{sec:conclusion}

We presented \methodname{}, a unified renderer for source-consistent and time-grounded multi-event audio generation. Its \tasg{} representation organizes persistent identities and local event requirements in a canonical format, while AbsTime supplies audio-side frame coordinates and valid duration. Joint attention and global modulation integrate these conditions, and shared-and-routed audio capacity supports heterogeneous mixtures of speech, music, ambience, and effects. Evaluations on established benchmarks and our SceneMix suite demonstrate improved compositional fidelity: \methodname{} achieves speech/sound F1 scores of 0.79/0.60 on MMAG and improves SceneMix event recall and window-hit rate over the matched natural-language variant by 10.8 and 15.9 percentage points, respectively. Controlled ablations support the contributions of structured conditioning, temporal grounding, and acoustic capacity, while human listening tests confirm improvements in audio quality and scene realization. These findings highlight the value of coupling organized scene specifications with time-grounded mixture rendering. Future work will address sub-second temporal control and explicit physical acoustic interactions among concurrent sources.

\bibliographystyle{IEEEtran}
\bibliography{ref}

@inproceedings{qiang2026unisonate,
  author={Qiang, C. and Wang, X. and Yin, K. and Liang, Y. and Guo, Y. and Ma, T. and Zhang, Z. and Wang, T. and Gong, C. and Chen, Y. and Fu, R. and Wang, L. and Dang, J.},
  title={{UniSonate}: A Unified Model for Speech, Music, and Sound Effect Generation With Text Instructions},
  booktitle={Proceedings of the Annual Meeting of the Association for Computational Linguistics (ACL)},
  pages={28043--28054},
  year={2026}
}

@article{mei2026dasheng,
  author={Mei, J. and Dinkel, H. and Niu, Y. and Sun, X. and Li, G. and Liao, Y. and Zhou, J. and Zhang, J. and Luan, J. and Wu, M.},
  title={{Dasheng AudioGen}: A Unified Model for Generating Coherent Audio Scenes From Text},
  journal={arXiv preprint arXiv:2605.27838},
  year={2026}
}

@article{cai2026sonicweave,
  author={Cai, Y. and Li, X. and Zhou, Y. and Li, J. and Wang, D. and Yang, D. and Wu, X. and Zhang, C. and Wu, Z. and Wan, P. and Meng, H.},
  title={{SonicWeave}: Chunk-Routed Mixture-of-Experts for Unified Audio Scene Generation},
  journal={arXiv preprint arXiv:2608.09571},
  year={2026}
}

@article{dai2026qwenaudio,
  author={Dai, J. and Duan, X. and Fan, X. and Feng, Y. and Li, J. and Li, X. and Li, Y. and Min, L. and Shi, Y. and Song, X. and Wang, Y. and Wen, C. and Wu, M. and Xiang, B. and Zhang, H. and Zhao, H. and Zheng, R.},
  title={{Qwen-Audio-3.0-Gen-Preview} Technical Report},
  journal={arXiv preprint arXiv:2607.27011},
  year={2026}
}

@article{sun2026midasheng,
  author={Sun, X. and Dinkel, H. and Li, G. and Mei, J. and Niu, Y. and Han, Z. and Jiang, Y. and Zhou, J. and Fan, L. and Luan, J.},
  title={{MiDashengLM-Gen}: Unified Audio Scene Generation via {LLM}-Driven Autoregressive Flow Matching},
  journal={arXiv preprint arXiv:2608.11804},
  year={2026}
}

@article{yang2026uniaudio2,
  author={Yang, D. and Wang, Y. and Chong, D. and Liu, S. and Wu, X. and Meng, H.},
  title={{UniAudio 2.0}: A Unified Audio Language Model With Text-Aligned Factorized Audio Tokenization},
  journal={arXiv preprint arXiv:2602.04683},
  year={2026}
}

@article{liu2026audiocalm,
  author={Liu, H. and Luo, K. and Wang, W. and Chen, Q. and Ma, B. and Li, X. and Xue, W.},
  title={{AudioCALM}: Continuous Autoregressive Language Modeling for Universal Audio Generation},
  journal={arXiv preprint arXiv:2606.23080},
  year={2026}
}

@article{lin2026stepaudio3,
  author={Lin, B. and Zhao, B. and Wang, B. and others},
  title={{StepAudio 3 Gen} Technical Report},
  journal={arXiv preprint arXiv:2609.12945},
  year={2026}
}

@article{huang2023makeanaudio2,
  author={Huang, J. and Ren, Y. and Huang, R. and Yang, D. and Ye, Z. and Zhang, C. and Liu, J. and Yin, X. and Ma, Z. and Zhao, Z.},
  title={{Make-An-Audio 2}: Temporal-Enhanced Text-to-Audio Generation},
  journal={arXiv preprint arXiv:2305.18474},
  year={2023}
}

@inproceedings{wang2025audiocomposer,
  author={Wang, Y. and Chen, H. and Yang, D. and Wu, Z. and Wu, X.},
  title={{AudioComposer}: Towards Fine-Grained Audio Generation With Natural Language Descriptions},
  booktitle={Proceedings of the IEEE International Conference on Acoustics, Speech, and Signal Processing (ICASSP)},
  year={2025}
}

@article{liu2026degdit,
  author={Liu, Y. and Li, C. and Zhang, W. and Wang, W. and Yu, M. and Fu, R. and Lin, Z. and Wang, W. and Yu, D.},
  title={{DegDiT}: Controllable Audio Generation With Dynamic Event Graph Guided Diffusion Transformer},
  journal={IEEE/ACM Transactions on Audio, Speech, and Language Processing},
  volume={34},
  pages={1300--1311},
  year={2026}
}

@inproceedings{evans2024timing,
  author={Evans, Z. and Carr, C. and Taylor, J. and Hawley, S. H. and Pons, J.},
  title={Fast Timing-Conditioned Latent Audio Diffusion},
  booktitle={Proceedings of the International Conference on Machine Learning (ICML)},
  pages={12652--12665},
  year={2024}
}

@article{zheng2025picoaudio2,
  author={Zheng, Z. and Xie, Z. and Xu, X. and Wu, W. and Zhang, C. and Wu, M.},
  title={{PicoAudio2}: Temporal Controllable Text-to-Audio Generation With Natural Language Description},
  journal={arXiv preprint arXiv:2509.00683},
  year={2025}
}

@inproceedings{jiang2026controlaudio,
  author={Jiang, Y. and Chen, Z. and Ju, Z. and Dai, Y. and Dou, W. and Zhu, J.},
  title={{ControlAudio}: Tackling Text-Guided, Timing-Indicated and Intelligible Audio Generation via Progressive Diffusion Modeling},
  booktitle={Proceedings of the Annual Meeting of the Association for Computational Linguistics (ACL)},
  pages={1394--1413},
  year={2026}
}

@inproceedings{jiang2025freeaudio,
  author={Jiang, Y. and Chen, Z. and Ju, Z. and Li, C. and Dou, W. and Zhu, J.},
  title={{FreeAudio}: Training-Free Timing Planning for Controllable Long-Form Text-to-Audio Generation},
  booktitle={Proceedings of the ACM International Conference on Multimedia (ACM MM)},
  year={2025}
}

@inproceedings{zhang2026tag,
  author={Zhang, H. and Zhang, J. and Zhang, C. and Wang, S. and Yang, W.},
  title={{TAG}: Structured Temporal Audio Generation via {LLM}-Guided Manual Scription and Control},
  booktitle={Proceedings of the IEEE International Conference on Acoustics, Speech, and Signal Processing (ICASSP)},
  year={2026}
}

@article{zhu2026audiocontrolnet,
  author={Zhu, H. and Xiao, Y. and Li, X. and Ma, Z. and Yu, J. and Zhang, B. and Yang, M. and Chen, X.},
  title={{Audio ControlNet} for Fine-Grained Audio Generation and Editing},
  journal={arXiv preprint arXiv:2602.04680},
  year={2026}
}

@article{xie2024audiotime,
  author={Xie, Z. and Xu, X. and Wu, Z. and Wu, M.},
  title={{AudioTime}: A Temporally-Aligned Audio-Text Benchmark Dataset},
  journal={arXiv preprint arXiv:2407.02857},
  year={2024}
}

@inproceedings{wang2025t2afeedback,
  author={Wang, Z. and Lei, K. and Zhu, C. and Huang, J. and Zhou, S. and Liu, L. and Cheng, X. and Ji, S. and Ye, Z. and Jin, T. and Zhao, Z.},
  title={{T2A-Feedback}: Improving Basic Capabilities of Text-to-Audio Generation via Fine-Grained {AI} Feedback},
  booktitle={Proceedings of the Annual Meeting of the Association for Computational Linguistics (ACL)},
  pages={23535--23547},
  year={2025}
}

@article{zheng2026mmag,
  author={Zheng, Z. and Xu, X. and Mei, J. and Li, Y. and Lv, M. and Wu, W. and Zhang, C. and Wu, M.},
  title={{MMAG}: A Multi-Control Mixed Audio Generation Benchmark},
  journal={arXiv preprint arXiv:2608.06900},
  year={2026}
}

@article{seedttseval2024,
  author={Anastassiou, P. and Chen, J. and Chen, J. and Chen, Y. and Chen, Z. and Chen, Z. and Cong, J. and Deng, L. and Ding, C. and Gao, L. and others},
  title={{Seed-TTS}: A Family of High-Quality Versatile Speech Generation Models},
  journal={arXiv preprint arXiv:2406.02430},
  year={2024}
}

@inproceedings{kim2019audiocaps,
  author={Kim, C. D. and Kim, B. and Lee, H. and Kim, G.},
  title={{AudioCaps}: Generating Captions for Audios in the Wild},
  booktitle={Proceedings of the Conference of the North American Chapter of the Association for Computational Linguistics: Human Language Technologies (NAACL-HLT)},
  pages={119--132},
  year={2019}
}

@article{agostinelli2023musiclm,
  author={Agostinelli, A. and Denk, T. I. and Borsos, Z. and Engel, J. and Verzetti, M. and Caillon, A. and Huang, Q. and Jansen, A. and Roberts, A. and Tagliasacchi, M. and others},
  title={{MusicLM}: Generating Music From Text},
  journal={arXiv preprint arXiv:2301.11325},
  year={2023}
}

@article{qwen2026tts,
  author={Hu, H. and Zhu, X. and He, T. and Guo, D. and Zhang, B. and Wang, X. and Guo, Z. and Jiang, Z. and Hao, H. and Guo, Z. and Zhang, X. and Zhang, P. and Yang, B. and Xu, J. and Zhou, J. and Lin, J.},
  title={{Qwen3-TTS} Technical Report},
  journal={arXiv preprint arXiv:2601.15621},
  year={2026}
}

@article{xu2025uniflow,
  author={Xu, X. and Mei, J. and Zheng, Z. and Tao, Y. and Xie, Z. and Zhang, Y. and Liu, H. and Wu, Y. and Yan, M. and Wu, W. and Zhang, C. and Wu, M.},
  title={{UniFlow-Audio}: Unified Flow Matching for Audio Generation From Omni-Modalities},
  journal={arXiv preprint arXiv:2509.24391},
  year={2025}
}

@article{chen2024f5tts,
  author={Chen, Y. and Niu, Z. and Ma, Z. and Deng, K. and Wang, C. and Zhao, J. and Yu, K. and Chen, X.},
  title={{F5-TTS}: A Fairytaler That Fakes Fluent and Faithful Speech With Flow Matching},
  journal={arXiv preprint arXiv:2410.06885},
  year={2024}
}

@article{du2025cosyvoice3,
  author={Du, Z. and Gao, C. and Wang, Y. and Yu, F. and Zhao, T. and others},
  title={{CosyVoice 3}: Towards In-the-Wild Speech Generation via Scaling-Up and Post-Training},
  journal={arXiv preprint arXiv:2505.17589},
  year={2025}
}

@misc{boson2025higgsv2,
  author={{Boson AI}},
  title={{Higgs Audio V2}: Redefining Expressiveness in Audio Generation},
  howpublished={Technical Report and Model Release},
  year={2025},
  url={https://github.com/boson-ai/higgs-audio}
}

@inproceedings{hung2026tangoflux,
  author={Hung, C.-Y. and Majumder, N. and Kong, Z. and Mehrish, A. and Zadeh, A. and Li, C. and Valle, R. and Catanzaro, B. and Poria, S.},
  title={{TangoFlux}: Super Fast and Faithful Text to Audio Generation With Flow Matching and {CLAP}-Ranked Preference Optimization},
  booktitle={Proceedings of the International Conference on Learning Representations (ICLR)},
  year={2026}
}

@inproceedings{copet2023musicgen,
  author={Copet, J. and Kreuk, F. and Gat, I. and Remez, T. and Kant, D. and Synnaeve, G. and Adi, Y. and D{\'e}fossez, A.},
  title={Simple and Controllable Music Generation},
  booktitle={Advances in Neural Information Processing Systems (NeurIPS)},
  volume={36},
  pages={47704--47720},
  year={2023}
}

@inproceedings{tian2026audiox,
  author={Tian, Z. and Liu, Z. and Jin, Y. and Yuan, R. and Xue, L. and Tan, X. and Chen, Q. and Xue, W. and Guo, Y.},
  title={{AudioX}: A Unified Framework for Anything-to-Audio Generation},
  booktitle={Proceedings of the International Conference on Learning Representations (ICLR)},
  year={2026}
}

@article{liu2024audioldm2,
  author={Liu, H. and Yuan, Y. and Liu, X. and Mei, X. and Kong, Q. and Tian, Q. and Wang, Y. and Wang, W. and Wang, Y. and Plumbley, M. D.},
  title={{AudioLDM 2}: Learning Holistic Audio Generation With Self-Supervised Pretraining},
  journal={IEEE/ACM Transactions on Audio, Speech, and Language Processing},
  volume={32},
  pages={2871--2883},
  year={2024}
}

@article{evans2024stableopen,
  author={Evans, Z. and Parker, J. D. and Carr, C. and Zukowski, Z. and Taylor, J. and Pons, J.},
  title={{Stable Audio Open}},
  journal={arXiv preprint arXiv:2407.14358},
  year={2024}
}

@inproceedings{liu2026unimoe,
  author={Liu, Z. and Li, Y. and Zhang, X. and Teng, Q. and Jiang, S. and Chen, X. and Shi, H. and Chen, H. and Meng, F. and Zhao, M. and others},
  title={{UniMoE-Audio}: Unified Speech and Music Generation With Dynamic-Capacity Mixture-of-Experts},
  booktitle={Proceedings of the Annual Meeting of the Association for Computational Linguistics (ACL)},
  pages={9107--9119},
  year={2026}
}

@misc{ming2026omnittts,
  author={{Inclusion AI}},
  title={{Ming-Omni-TTS}: Simple and Efficient Unified Generation of Speech, Music, and Sound With Precise Control},
  howpublished={Project Report and Model Release},
  year={2026},
  url={https://github.com/inclusionAI/Ming-omni-tts}
}

@article{ren2026audirector,
  author={Ren, Y. and Xu, X. and Zhang, Z. and Wu, W. and Li, B. and Zhang, C.},
  title={{AuDirector}: A Self-Reflective Closed-Loop Framework for Immersive Audio Storytelling},
  journal={arXiv preprint arXiv:2605.11866},
  year={2026}
}

@article{hacohen2026ltx2,
  author={HaCohen, Y. and Brazowski, B. and Chiprut, N. and Bitterman, Y. and Kvochko, A. and others},
  title={{LTX-2}: Efficient Joint Audio-Visual Foundation Model},
  journal={arXiv preprint arXiv:2601.03233},
  year={2026}
}

@inproceedings{zhang2026uniavgen,
  author={Zhang, G. and Zhou, Z. and Hu, T. and Peng, Z. and Zhang, Y. and Chen, Y. and Zhou, Y. and Lu, Q. and Wang, L.},
  title={{UniAVGen}: Unified Audio and Video Generation With Asymmetric Cross-Modal Interactions},
  booktitle={Proceedings of the IEEE/CVF Conference on Computer Vision and Pattern Recognition (CVPR)},
  year={2026}
}

@inproceedings{esser2024mmdit,
  author={Esser, P. and Kulal, S. and Blattmann, A. and Entezari, R. and M{\"u}ller, J. and Saini, H. and Levi, Y. and Lorenz, D. and Sauer, A. and Boesel, F. and Podell, D. and Dockhorn, T. and English, Z. and Rombach, R.},
  title={Scaling Rectified Flow Transformers for High-Resolution Image Synthesis},
  booktitle={Proceedings of the International Conference on Machine Learning (ICML)},
  pages={12606--12633},
  year={2024}
}

@inproceedings{lipman2023flow,
  author={Lipman, Y. and Chen, R. T. Q. and Ben-Hamu, H. and Nickel, M. and Le, M.},
  title={Flow Matching for Generative Modeling},
  booktitle={Proceedings of the International Conference on Learning Representations (ICLR)},
  year={2023}
}

@inproceedings{sadat2025apg,
  author={Sadat, S. and Hilliges, O. and Weber, R. M.},
  title={Eliminating Oversaturation and Artifacts of High Guidance Scales in Diffusion Models},
  booktitle={Proceedings of the International Conference on Learning Representations (ICLR)},
  year={2025}
}

@article{qwen3-vl,
  author={Bai, S. and Cai, Y. and Chen, R. and others},
  title={{Qwen3-VL} Technical Report},
  journal={arXiv preprint arXiv:2511.21631},
  year={2025}
}

@article{qwen2vl,
  author={Wang, P. and Bai, S. and Tan, S. and Wang, S. and Fan, Z. and Bai, J. and Chen, K. and Liu, X. and Wang, J. and Ge, W. and Fan, Y. and Dang, K. and Du, M. and Ren, X. and Men, R. and Liu, D. and Zhou, C. and Zhou, J. and Lin, J.},
  title={{Qwen2-VL}: Enhancing Vision-Language Model's Perception of the World at Any Resolution},
  journal={arXiv preprint arXiv:2409.12191},
  year={2024}
}

@article{liu2025wavjourney,
  author={Liu, X. and Zhu, Z. and Liu, H. and Yuan, Y. and Huang, Q. and Cui, M. and Liang, J. and Cao, Y. and Kong, Q. and Plumbley, M. D. and Wang, W.},
  title={{WavJourney}: Compositional Audio Creation With Large Language Models},
  journal={IEEE/ACM Transactions on Audio, Speech, and Language Processing},
  volume={33},
  pages={2830--2844},
  year={2025}
}

@article{duan2026audiooscar,
  author={Duan, Y. and Xu, Q. and Wu, H. and Liu, Z. and Guan, W. and Liu, J. and Ma, Z. and Xu, K. and Chen, X.},
  title={{Audio-Oscar}: A Multi-Agent System for Complex Audio Scene Generation, Orchestration, and Refinement},
  journal={arXiv preprint arXiv:2606.07397},
  year={2026}
}

@article{finkelson2026scena,
  author={Finkelson, M. and Segal, D. and Richardson, E. and Armon, S. and Goldring, N. and Panet, P. and Zabari, N. and Brazowski, B. and Patashnik, O. and HaCohen, Y.},
  title={Reference-Driven Multi-Speaker Audio Scene Generation From In-the-Wild Priors},
  journal={arXiv preprint arXiv:2606.19325},
  year={2026}
}

@inproceedings{peng2026vibevoice,
  author={Peng, Z. and Yu, J. and Wang, W. and Chang, Y. and Sun, Y. and Dong, L. and Zhu, Y. and Xu, W. and Bao, H. and Wang, Z. and Huang, S. and Xia, Y. and Wei, F.},
  title={{VibeVoice}: Expressive Podcast Generation With Next-Token Diffusion},
  booktitle={Proceedings of the International Conference on Learning Representations (ICLR)},
  year={2026}
}

@article{xie2024picoaudio,
  author={Xie, Z. and Xu, X. and Wu, Z. and Wu, M.},
  title={{PicoAudio}: Enabling Precise Timestamp and Frequency Controllability of Audio Events in Text-to-Audio Generation},
  journal={arXiv preprint arXiv:2407.02869},
  year={2024}
}

@inproceedings{guo2024mcdiffusion,
  author={Guo, Z. and Mao, J. and Tao, R. and Yan, L. and Ouchi, K. and Liu, H. and Wang, X.},
  title={Audio Generation With Multiple Conditional Diffusion Model},
  booktitle={Proceedings of the AAAI Conference on Artificial Intelligence},
  volume={38},
  number={16},
  pages={18153--18161},
  year={2024}
}

@inproceedings{du2024tgdiff,
  author={Du, T. and Chen, J. and Lu, J. and Xu, Q. and Liao, H. and Chen, Y. and Wu, Z.},
  title={Controllable Text-to-Audio Generation With Training-Free Temporal Guidance Diffusion},
  booktitle={Proceedings of the IEEE International Conference on Multimedia and Expo (ICME)},
  year={2024}
}

@inproceedings{kuan2026instruction,
  author={Kuan, C.-Y. and Kim, S. and Kim, B. and Kim, S. and Lu, B.-R. and Tang, Q. and Gandhe, A. and Lee, H.-Y. and Kao, C.-C. and Wang, C.},
  title={Improving Text-to-Audio Instruction Following via Fine-Grained Feedback From Audio-Aware Large Language Models},
  booktitle={Proceedings of Interspeech},
  year={2026}
}

@article{liu2026temporallink,
  title={The Missing Temporal Link: Temporal Context Routing for Script-Driven Audio-Video Generation},
  author={Liu, Yichen and Zhang, Quanwei and Wang, Haozhe and Zhou, Donghao and Zhang, Jiankun and Li, Xiaojie and Shi, Yang and Liu, Jiaming and Huang, Ruihua and Zou, Yingtian and others},
  journal={arXiv preprint arXiv:2609.02367},
  year={2026}
}

@article{rong2026audiomap,
  author={Rong, Y. and Ma, F. and Li, X. and Wang, J. and Zhang, C. and Liu, L.},
  title={{AudioMap}: Cloze-and-Choice Reinforcement Learning for Time-Aware Dense Audio Captioning},
  journal={arXiv preprint arXiv:2608.09559},
  year={2026}
}

@article{fedus2022switch,
  author={Fedus, W. and Zoph, B. and Shazeer, N.},
  title={{Switch Transformers}: Scaling to Trillion Parameter Models With Simple and Efficient Sparsity},
  journal={Journal of Machine Learning Research},
  volume={23},
  number={120},
  pages={1--39},
  year={2022}
}

@article{liao2026fishaudios2,
  title   = {{Fish Audio S2} Technical Report},
  author  = {Shijia Liao and Yuxuan Wang and Songting Liu and
             Yifan Cheng and Ruoyi Zhang and Tianyu Li and
             Shidong Li and Yisheng Zheng and Xingwei Liu and
             Qingzheng Wang and Zhizhuo Zhou and Jiahua Liu and
             Xin Chen and Dawei Han},
  journal = {arXiv preprint arXiv:2603.08823},
  year    = {2026}
}

@article{zhou2026voxcpm2,
  title   = {{VoxCPM2} Technical Report},
  author  = {Yixuan Zhou and Guoyang Zeng and Xin Liu and
             Xiang Li and Renjie Yu and Jiancheng Gui and
             Jiaheng Wu and Ziyang Wang and Xudong Shen and
             Runchuan Ye and Zhisheng Zhang and Jiuyang Zhou and
             Bingsong Bai and Weiyue Sun and Mengyuan Deng and
             Qundong Shi and Zhiyong Wu and Zhiyuan Liu},
  journal = {arXiv preprint arXiv:2606.06928},
  year    = {2026}
}
\end{document}